\documentstyle[12pt,aaspp4]{article}

\received{}
\accepted{}
\slugcomment{To Appear in the Astrophysical Journal}

\begin{document}
\include{psfig}
\def\kms{km~s$^{-1}$ }
\def\Lya{Ly$\alpha$ }
\def\lya{Ly$\alpha$ }
\def\Lyb{Ly$\beta$ }
\def\lyb{Ly$\beta$ }
\def\Lyg{Ly$\gamma$ }
\def\Lyd{Ly$\delta$ }
\def\Lye{Ly$\epsilon$ }
\def\d{$d_5$ }
\def\Ly{Lyman}
\def\ang{\AA }
\def\gq{$\geq$ }
\def\zem{$z_{em}$ }
\def\zabs{$z_{abs}$ }
\def\cm2{cm$^{-2}$ }
\def\nht{N(H~I)$_{total}$ }
\def\cmin{$\chi^2_{min}$ }

%
\title{The Deuterium Abundance Towards Q1937--1009}

\altaffiltext{1}{Visiting Astronomer, W. M. Keck Telescope, California
Association for Research in Astronomy}
 
\altaffiltext{2}{Present Address: University of Chicago, 5640 S. Ellis Ave.,
Chicago, IL 60637}

\altaffiltext{3}{Visiting Associate, European Southern Observatory.}
 
%
%

\author{Scott Burles\altaffilmark{1,2} \& David Tytler\altaffilmark{1,3}} 
\affil{Department of Physics, and Center for Astrophysics and Space
Sciences \\
University of California, San
Diego \\
C0424, La Jolla, CA 92093-0424}
 
\begin{abstract}

We present a new measurement of the deuterium-to-hydrogen
ratio (D/H) in the Lyman limit absorption system at $z = 3.572$ towards
Q1937--1009.  Tytler, Fan \& Burles (1996; hereafter TFB) 
made the first extragalactic detection of deuterium in this absorption 
system, which remains the best location for a high accuracy measurement 
of primordial D/H.  Their detailed analysis of Keck spectra 
gave a low value of D/H = 
$2.3 \pm 0.3 \pm 0.3 \times 10^{-5}$ ($1\sigma$ statistical \& systematic
errors).  Now we present a new method to measure D/H in QSO
absorption systems. We avoid many of the assumptions adopted by TFB; 
we allow extra parameters to
treat the continuum uncertainties, 
include a variety of new absorption models
which allow for undetected velocity structure,
and use the improved measurement of the 
total hydrogen column density by Burles \& Tytler
(1997a).  We find that all models, including contamination, give an upper limit
D/H $< 3.9 \times 10^{-5}$ (95 \% confidence).  Both this and previous
analyses find contamination to be unlikely in this absorption system,
A $\chi^2$ analysis in models without contamination gives
D/H $= 3.3 \pm 0.3 \times 10^{-5}$ (67\% confidence), 
which is higher but consistent with the earlier results of TFB, and a second
measurement of D/H towards Q1009+2956 (Tytler \& Burles 1997).  
With calculations of standard big bang nucleosynthesis (SBBN) and the
assumption that
this measurement of D/H is representative of the primordial value, 
we find a high baryon to photon ratio, 
$\eta = 5.3 \pm 0.4 \times 10^{-10}$.
This is consistent with primordial abundance determinations
of $^4$He in H~II regions
(Izotov et al. 1997) and $^7$Li in the atmospheres of 
warm metal-poor population II stars (\cite{bon97}).
We find a high value for the present-day baryon density,
$\Omega_b \, h^2 = 0.0193 \pm 0.0014$, which is consistent 
with other
inventories of baryonic matter, from low to high redshift:
clusters of galaxies, the Lyman alpha forest \& the Cosmic Microwave 
Background.

\end{abstract}

\section{INTRODUCTION}

Standard Big Bang Nucleosynthesis
(SBBN; \cite{wag67}; \cite{wal91};~\cite{smi93}; \cite{kra95}; \cite{cop95})
predicts that a measurement of the primordial ratio D/H will give the most
sensitive constraint on the cosmological baryon to photon ratio, $\eta$.
Using the photon density measured from the temperature of the Cosmic
Microwave Background (CMB), 
we can then obtain the most
sensitive constraint on $\Omega_b h^2$, the cosmological baryon density 
in units of the critical density,
where $H_0 = 100 \, h$ \kms Mpc$^{-1}$.
Both $\eta$ and $\Omega_b$ are fundamental cosmological parameters,
and measurements of D/H in QSO absorption systems can determine
both to 10\% accuracy.  
When we compare measurements of the primordial abundances of 
different light elements,
H, D and $^4$He in particular, we test the SBBN theory predictions.
Measurements of D/H in different astrophysical sites also constrains the 
history of star formation in these sites, once the primordial value of D/H 
is established, because D is totally destroyed (astrated) as gas is cycled
through and ejected from stars (\cite{eps76}).
 
Adams (1976) first suggested that deuterium could be measured in
intergalactic gas clouds through Lyman absorption lines 
in the spectra of distant QSOs. 
Deuterium Lyman absorption lines are 82 \kms on the
short wavelength side of the
corresponding hydrogen Lyman lines, but Adams notes that D can be detected
in only very select absorption systems with simple velocity structure
and high neutral hydrogen column density,
N(H~I) $> 10^{17}$ \cm2 (Lyman limit systems).
For high redshifts, $z > 2.5$, the Lyman series
is redshifted into the optical, and the lines are accessible to
large ground based telescopes.
The Lyman absorption lines of D and H, and the 
Lyman continuum absorption of H~I, constrain the column densities 
of D~I and H~I, and provide a measurement of D~I/H~I.  
The timescale for H and D ionization equilibrium
is very short ($10^5$ yr) in the photoionized intergalactic medium
($n_H \approx 10^{-3}$), and we assume the ratio D~I/H~I is identical
to D/H throughout this paper (\cite{ike86}).

The Lyman limit system at $z=3.572$ towards Q1937--1009 is 
an ideal site to infer a primordial value for D/H.  The system
is very metal-poor, less than 1/100 solar.  Deuterium is destroyed
as gas is cycled through stars, but metals are produced in the cycle
and the system's low metallicity limits the amount of deuterium which
could have been destroyed.  The gas in the system also has very 
low internal velocities, which can limit the amount of kinetic 
energy input through any high energy phenomena, such as supernovae or
gamma ray sources.  

In TFB, we presented
high resolution spectra of Q1937--1009, and made the first
measurement of D/H in a QSO absorption system.
This system has a high total neutral hydrogen column density, 
Log N(H~I) $= 17.86 \pm 0.02$ \cm2 (Burles \& Tytler 1997a), 
and the corresponding absorption is optically thick throughout the
entire Lyman series.  The deuterium feature is well determined
by its profile in Ly$\alpha$, and the column density of D
is well constrained.

In this paper, we advance the methods of TFB to place more robust
constraints on D/H in this absorption systems.
In section 2, we describe our new method of constraining D/H, and
in section 3, we discuss the new results.
In section 4, we calculate the abundances for a number of metals
and discuss the ionization state of the system.  In section 5, we 
compare our D/H value to the abundance of D in other sites,
and other light elements, and we show that our D/H is probably
primordial.

\section{CONSTRAINING D/H}

We model the absorption spectrum along the QSO line of sight as a finite number
of discrete absorbing components.  Each component is modeled as
a Voigt profile (Spitzer 1979),
given by 
three parameters: column density (N), redshift ($z$), 
and velocity dispersion ($b$).  
At every wavelength, we sum the optical depth contribution of all
absorption line profiles in the model.  
The optical depth is converted to a normalized flux,
and convolved with the instrumental response.  The model spectrum can then
be directly compared to the observed spectrum.  We use the
observed spectrum described in TFB.

\subsection{Previous Analysis}

The TFB analysis was sophisticated since it was the first to
simultaneously fit many metal and H lines, and to separate thermal and
turbulent contributions to line widths. However, it also used many
assumptions which we now remove.

In TFB, 
we modeled the absorption system with two D/H components, labeled
``blue" and ``red".  The strong \Lya feature also required a third H~I 
component (which contained only 1\% of the hydrogen column) to obtain
an adequate fit to the spectrum of the Lyman series.  The positions of
the two main
components were ``tied" to the velocity positions determined from the
narrow metal lines which were asymmetric, and well described by these
two components.  We fit the model to narrow
regions of the spectrum
containing only the Lyman lines of interest, and did not include
absorption from other \Lya lines along the line of sight.  D/H was
assumed to be equal in both components.  The velocity dispersion of the
red D line was determined from the dispersions of the corresponding
H and metal lines in the red component.  We found the value of D/H which
gave the best fit to the data and the formal $1\sigma$ errors
from the D and H column densities, log(D/H) = --4.64 $\pm$ 0.06.  We
investigated systematic errors due to incorrect continuum placement in
the regions of \Lya and the Lyman limit, and estimated the
systematic error $\Delta$ log(D/H) = 0.06.

The TFB analysis found 
a consistent model and a well-defined value for the most likely value
of D/H.  But its limitations lay in the assumptions and
the calculation of confidence levels
and the total errors on D/H.

\subsection{Present Analysis}

In this paper, we present major improvements in our analytic
techniques.  We measure D/H with a variety of models and 
free parameters.  We add new free parameters to describe the unabsorbed   
quasar spectrum, which we use to normalize the absorption spectrum.
Previously, we treated
continuum uncertainties as systematic errors. 
The number and velocities of the D/H components are now also 
free parameters.  We no longer require that the
D and H lines have the same velocities as the metal lines.  The D and H
line profiles are 
constrained only by the spectra covering the Lyman series lines.
To include all the uncertainties from the modeling into the final uncertainty
of D/H, we perform a $\chi^2$ analysis as a function of D/H.  Formerly, we
calculated the errors from a quadrature sum of the total column density 
uncertainties of D and H.  By calculating $\chi^2$ as a function of D/H,
we can include the uncertainties from all the free parameters in our models,
including the velocity dispersions, velocity positions, continuum levels,
and column densities.  The $\chi^2$ function
gives the relative likelihood of all values of D/H versus the 
best fit value.  The final confidence region 
can be directly calculated from the $\chi^2$ function,
and can be considered more ``comprehensive" than the uncertainties 
which we reported before.

\subsubsection{Total Hydrogen Column Density}

The most critical addition to the present analysis is a measurement of
the total hydrogen column density, \nht. From our detailed study of the
Lyman continuum optical depth (Burles \& Tytler 1997a), we measured
\nht = 17.86 $\pm$ 0.02,
and we use this constraint for all models.
\nht is much better constrained
from the Lyman continuum absorption than from the
Lyman series line profiles, which gave 
\nht = 17.94 $\pm 0.06 \pm 0.06$ (TFB), 
where the first error is statistical
and the second is systematic.  The new \nht provides a very strong
constraint for the models, allows us to add more free parameters 
to the models to test a variety of models and minimize the model
fit with respect to D/H.

\subsubsection{New Models}

Here we discuss the new models we have adopted to measure D/H.
The absorbers in each model can be 
placed into one of two groups, depending on their
association with the D/H absorption system (DHAS).   
The first group contains H~I absorbers which will not 
show D~I (also labeled ``unassociated"), 
and the second group contains the high column H~I
absorbers which show D~I and are associated with the DHAS.  We refer to
absorbers in the second group as the ``main components''.
For each unassociated absorber, we add three free parameters to the model:
one each for N, $b$, and $z$.  
Each main component shows two lines: one of H~I and the other D~I.
We also add one parameter for each line to allow a separation of $b$ into
T and $b_{tur}$, because in general $b$(H~I) can differ from $b$(D~I),
but we tie three of the parameters in the two lines:
$z\rm{(H~I)} = z\rm{(D~I)}$, T(H~I) = T(D~I), and $b_{tur}$(H~I) = 
$b_{tur}$(D~I).  
In this paper, we do not use the metal lines to constrain D/H. 
The velocity dispersions of the main components are given by
$b^2 = 163.84 (T / m)+ b^2_{tur}$, where $T$ is temperature
in units of $10^4$ K, $m$ is the atomic mass, and $b$ is in units of \kms.
For zero turbulent velocity dispersion,
we recover the thermal relation for the H~I and D~I lines,
$b\rm{(H~I)} = \sqrt{2} \, b\rm{(D~I)}$.  
In all models we require that $T >$ 5000 K in the main components.
Without this constraint, models which fit the data 
may include unobserved components with very low T. The background ionizing 
radiation is hard enough that the absorbing gas is unlikely to be cooler than 
a few 10$^4$ K. 

The results depend on assumptions we make when constructing
the models.  These assumptions include the number of D/H components,
the total number of free parameters, and the spectral regions which we
fit. 
The total number of free parameters is the sum of parameters of all
the absorption components and the free parameters included in the
continuum.  For this analysis, we have chosen 7 models, 
and we measure D/H for each.
We can then compare the best fit values of D/H
for each model, and assess the effect of the different
assumptions used in each model.
All 7 models include the spectral regions listed in Table 1, 
the number of D/H components chosen for each model, and
the 64 H~I absorbers listed in Table 2.  

The models are of three types. Models 1 \& 2
are the simplest and include 2 and 3 D/H components, respectively,
with no free parameters in the continuum.  The continuum is held
fixed at the initial placement 
of the unabsorbed continuum level, which is shown as the solid lines
at unity in Figure 2.
The other five models allow for free parameters in the continuum
in each region.  The number of allowed continuum 
free parameters in each region is shown in Table 2.
The final two models have three D/H components, but also allow
for hydrogen contamination.  In Model 6, an extra hydrogen absorber
is introduced at redshift $z=3.570958$, which places H~I Ly$\alpha$ at 
the position of D-Ly$\alpha$, and H~I Ly$\beta$ at D-Ly$\beta$.
In Model 7, the extra hydrogen absorber is introduced at $z=2.85670$, 
which places its \Lya at 
the position of D-Ly$\beta$, in the blue wing of \Lyb.  
The model parameters are summarized in
Table 3.

\subsubsection{Fitting Procedure}

We use the Levenberg-Marquardt method to 
minimize $\chi^2$ (Press et al. 1992).  The details of the algorithm and
computational techniques are 
presented in Burles (1997).
The algorithm iterates until it converges on the best
fit model.  The $\chi^2$ of the final fit is assumed to be the minimum
$\chi^2$, $\chi^2_{min}$.

A given model will have $M$ free parameters, and the algorithm calculates
a $M \times M$ covariance matrix for each iteration.
The 1$\sigma$ uncertainties can be directly calculated from the 
diagonal elements in the covariance matrix of the final iteration.

In all the models, we are interested in one parameter specifically,
D/H.  It is not straight forward to calculate the total uncertainty
in D/H from the final covariance matrix directly, even if the
errors are normally distributed.  Due to the intrinsic blending
of the H~I and D~I lines, the formal errors in column density are
correlated. 
Instead, we choose
to make D/H a free parameter in the models, and calculate \cmin
as a function of the parameter D/H.

We construct a list of D/H values, and model the spectrum for each value
in the list.  By designating D/H in each model, we reduce the
number of free parameters, $M$, by one for each D/H component in the
model.  Now each main component (components which have D absorption)
has only four free parameters: N(H~I), $T$, $b_{tur}$, and $z$.
N(D~I) is no longer free, but is given by N(D~I) = N(H~I) $\times$ (D/H). 
We assume that D/H is the same in all main components.

By constructing an array of \cmin as of function of D/H, we can 
calculate the most likely value of D/H and the confidence levels
surrounding this value.  We have effectively taken a cross section
of the $M$-dimensional $\chi^2$ function
to calculate the uncertainties of the single
parameter, D/H.  The one-dimensional function, $\chi^2\rm{(D/H)}$,
yields both the most likely value of D/H, and confidence levels about
the most likely value.

\subsubsection{Continuum Level}

Figure 1 shows the spectral regions stacked in velocity space.  
Figure 2 presents each spectral region separately on the vacuum heliocentric
wavelength scale.  

The unabsorbed QSO continuum is now allowed to vary to achieve the best fit.
The QSO continuum is modeled by a low-order Legendre polynomial, which
accounts
for smooth variations in the continuum on scales of 100 \kms (dashed lines
in Fig. 2).
In the data reduction process, we use Legendre polynomials to 
initially normalize the spectrum 
to the regions of the spectrum showing no absorption (solid line at unity in
Fig. 2).
In the \Lya forest, the continuum is not well defined due to large regions
of continuous absorption.  We account for this uncertainty in the continuum
in the \Lya forest by allowing the coefficients of the Legendre polynomials
to be free parameters, without bounds, in our models.
Therefore, the continuum is no longer
fixed to the level on which we can only speculate.  
In statistical language, we fit the continuum level with a set
of ``nuisance parameters", which we allow to vary freely to improve
the fit, because they are not of primary interest.

\subsubsection{Regions of Interest}

We must choose regions of the spectrum to compare to the models.
If we model the entire Keck spectrum, the model would include
over 1000 lines and over 40,000 pixels.  For practical purposes, we can
not model the entire spectrum simultaneously, the time required
to complete the fitting procedure scales as the number of lines times 
the number of pixels.  Also, we want our model to be sensitive to D/H,
and not other unrelated features in the spectrum.  Therefore, we select
regions of the spectrum which include
at least one Lyman series line of the DHAS.  In principle, we want to 
include all Lyman series of the DHAS, but some Lyman
series lines are blended with other strong, unassociated absorption
lines, and are not included in the selected regions.   

In Figures 1 \& 2, we show the regions of the spectrum
used in the model fitting.  Table 1 lists the regions used in our analysis,
including the number of pixels and the order of the Legendre
polynomial used to model the unabsorbed continuum in each region.
The damping wings of the \Lya feature absorb over a large wavelength
range, therefore the span of the \Lya region is much greater than the
other individual line regions.
For the higher-order Lyman lines, the regions of interest begin to
overlap.   The overlapping regions are combined into a single region
containing multiple Lyman lines and this region is labeled 
``Ly-limit".

\subsubsection{Goodness of Fit to Different Spectral Regions}

Of the 9 regions, the \Lya region (Figure 2a) has the highest SNR per pixel, 
approximately 75 per 4 \kms pixel.  Thus the entire fit is heavily weighted 
by Ly$\alpha$, and the fit is most tightly constrained by 
the data is in this region.  Deuterium \Lya
can be seen at 5557 \AA, and is composed of 3 components in the model presented
here.  The fitted continuum contains 5 free parameters, and is reasonably
consistent with our initial estimate of the continuum.  Only at the red edge
does it begin to diverge significantly.  And the overlap between the estimated
and the best fit continua lies directly at the \Lya feature.  The overlap
in this region suggests that the new continuum determination will have little
effect in the \Lya region.

The \Lyb region (Figure 2b) is much smaller than Ly$\alpha$.  The flux returns 
to the continuum much closer to the center of the Ly$\beta$ line, 
and there is no reason to
include a larger section around Ly$\beta$.  Again, the best fit continuum lies
close to the original continuum estimate over the entire region.  
The model fit to \Lyb is not as good as Ly$\alpha$, there is  
under-absorption in two places near D-\Lyb.  The blue wing of \Lyb shows
under-absorption, but there is a significant increase in the 
noise
over some of these pixels ($4689.0 < \lambda < 4689.3$ \AA), which is most
likely due to bad columns being rejected in the original CCD images.
The under-absorption on the blueward side of D-\Lyb is likely due to 
a hydrogen \Lya line (which is fit in Model 7)  
that is blended with the deuterium feature.  We 
label hydrogen lines which overlap and blend with the deuterium features
``contaminating" hydrogen.  Contamination will be discussed thoroughly in
section 3.1, and introduced into our Models 6 \& 7. 

The models must produce a good fit to all the Lyman lines simultaneously.
We show each of the regions separately in Fig. 2 
to allow a close inspection of the data and model, and to 
display how well the model reproduces the observed spectrum.
Notice that there are additional absorbers
overlapping the main components. These extra absorbers add free
parameters to the fit, and in general, make the model less restrictive.
That is, the parameters determining the main components cannot be as tightly
constrained when overlapping absorbers are included.  

In Figure 2i, we show the spectral region with the highest order Lyman
lines, Ly-12 to Ly-19.  The best fit continuum shows a significant difference
from the initial estimate.  The unabsorbed continuum in this region was
difficult to estimate due to the lack of pixels with little or no absorption.
The continuum was originally estimated as a constant flux level
passing near the few pixels with the highest flux. The large amount of 
absorption in this region did not allow for a better estimate of the unabsorbed
continuum initially.

The difference between the best fit continuum (dotted line) and the initial
continuum (solid line) is likely due to the simple approximation of the 
continuum level during the reduction.  
The shape of the best fit continuum is more likely 
and is a function of the instrumental sensitivity in this region.

There are several regions of the spectrum shown in
Figure 2 of TFB which have more flux than expected by their model fit.
In the models presented here, this unabsorbed flux is accounted for 
with the extra hydrogen components and a different continuum.

\section{RESULTS}

Figure 3 shows the major results of the fitting procedure with the
seven 
models.  Table 3 summarizes the D/H measurements 
for each model.  We performed the 
$\chi^2$ minimization procedure for 100 values of D/H over the
range $-4.95 <$ Log(D/H) $< -4.0$ for each of the seven models. 

For example, Figure 3a shows the results for the 2 component
model (Model 1).  The solid dark gray line represents one of the components
and the dashed line represents the other.
These graphs show the behavior of the main component parameters as
a function of D/H.  The parameters sometimes display discontinuous behavior,
which is due to the non-uniqueness of Voigt profile fitting.  The $\chi^2$
minimization has found two equally likely solutions, and this exhibits
itself with a discontinuous change in the parameter values.
Although the fitted parameters may be non-unique, \cmin remains 
a smooth function of D/H.  The exact parameters which give the
best fit are not well-determined, but the \cmin, and the D/H are
well defined.
In this analysis, we cannot attempt to model the DHAS exactly, but 
we can still find the relative likelihood for all values of D/H.

We cannot model the individual components exactly because of the 
intrinsic blending of the absorption lines.  But we do find that parameters
which include all the components can be measured precisely.  In Figure 3,
the thick solid line in the column density plot represents the 
total neutral hydrogen column, \nht.  This is merely the sum of the
column densities of all components in a given model at a specific value
of D/H.  We find that all seven models give a \nht which is a smooth
function of D/H, even though the individual column densities in the
models are not.  \nht is the quantity subject to the constraint 
provided by Burles \& Tytler (1997a), \nht = 17.86 $\pm$ 0.02.  For a
model which gives an arbitrarily good fit to the spectral regions,
the best fit will depend only on the agreement with the \nht constraint.

We can measure D/H directly from Figure 3.  With a 
$\chi^2$ function of one variable, the confidence levels are easily
determined.  The 95\% confidence levels correspond to $\Delta \, \chi^2
= 4.0$, where $\Delta \, \chi^2 = \chi^2 -$ \cmin.  For each model
in Figure 3, we find the minimum of the $\chi^2$ function and list
this value in Table 3.  The 95\% confidence levels on D/H are also listed in
Table 3, and are shown as vertical dashed lines in Figure 3.
The fitting procedure is not perfect, and we can see in Figure 3
that there are deviations from perfectly smooth $\chi^2$ functions.

The fitting procedure can converge before reaching the absolute minimum
$\chi^2$, and we estimate that the difference between the absolute
minimum and the calculated $\chi^2$ can be as large as 0.5.
We take this deficiency into account by setting the 95\% confidence
levels at $\Delta \, \chi^2 = 4.5$.  

All of the regions of 95\% confidence in Figure 3 overlap with the
central value of \nht constraint.  The best fit model
of D/H must not only provide the best fit to the spectral regions, but
must also fit the constraint of \nht.  For all models, 
the values of \nht decrease smoothly
with increasing D/H.  
N(D~I) is tightly constrained by D-\Lya and D-\Lyb, so as D/H 
increases, N(H~I) must decrease, which  
gives smoothly
varying functions of \nht for each model. 
In all models, the 95\% confidence
regions on D/H represent the model fits where the
\nht constraint is well satisfied.

All seven models give consistent
ranges for the 95\% confidence regions, as seen in Figure 4. 
As expected, the models with
more free parameters have larger confidence regions.  Except for Model
6, which includes hydrogen contamination at Ly$\alpha$, 
all models are consistent with 
\begin{equation}
\rm{Log (D/H)} = -4.49 \pm 0.04,
\end{equation}
or
\begin{equation}
\rm{(D/H)} = 3.24 \pm 0.30 \times 10^{-5},
\end{equation}
at 67\% confidence.
The shape of the
\cmin functions 
indicate errors which are normally distributed in Log (D/H), and the
67\% confidence levels should be one-half of the 95\% levels.

\subsection{Contamination}

In Models 6 \& 7, we investigate the effects of hydrogen 
contamination of the D-Lyman lines.  Model 6 includes an additional
hydrogen absorber at redshift, $z=3.5710$, which places its
lines near the deuterium lines.  Figure 3f shows the results of
the fitting procedure.  For values of Log D/H $> -4.5$, the results
are identical to Model 4, which did not include contamination.
But for lower values of D/H, the contaminating hydrogen absorbs a
significant amount of flux at the D-\Lya and D-\Lyb lines,
and gives a good fit by keeping \nht nearly constant and lowering
N(D~I).  Although contamination allows for lower D/H as expected,
the introduction of an additional \Lya absorber with three free parameters
does not improve the fit significantly.  This result is in sharp contrast
to another D/H system, towards Q1009+2956, where contamination significantly
improves $\chi^2$ (Burles \& Tytler 1997b).  We conclude that \Lya contamination is not significant
in this D/H system for two reasons: (1) additional parameters for contamination
do not improve the $\chi^2$ fit, (2) the likelihood of significant
contamination drawn from distributions of the \Lya forest (Kirkman \& Tytler
1997; Lu et al. 1997) is small (c.f. Tytler \& Burles 1997, TFB, 
\cite{jed97b}, Steigman 1994).

Model 7 includes contamination at the position of D-\Lyb, 
at $z=2.8565$.  This redshift is much lower than the DHAS, and the
hydrogen absorber only affects D-\Lyb.
The extra absorber is initially placed at the center of D-\Lyb, and
is allowed to freely move to achieve the best fit to the data.
The fit improves by
$\Delta \chi^2 \approx 7.0$, as the hydrogen absorber fills
in the under-absorption seen in Figure 2b.  The best fit for Model 7
is shown in Figure 5, and the improvement is easily seen upon
comparison with Figure 2b.  The model shows under absorption in three pixels
near 4689 \AA (-60 km/s blueward of H-\Lyb), but these pixels fall in a region
of spectra with increased noise.
In conclusion, we find Model 7 is 
consistent with the results from the other models
which did not include this contamination and the 
measurement of D/H is robust with any extra hydrogen absorption at \Lyb alone.

The presence of contamination will give an overestimate of D/H.  Therefore,
the upper limit on D/H is robust and not 
affected by the presence of contamination.
All seven models are consistent with the upper limit, 
D/H $< 3.9 \times 10^{-5}$.

\section{Ionization and Metals}

The metal absorption lines 
associated with the DHAS were analyzed by TFB.
We use the two component fit ($z = 3.572201, 3.572428$) of TFB
to measure the column densities of the
ions shown in Table 4.  If no feature is detected at the expected positions
of the ionic transitions, then we place a 2$\sigma$ upper limit on the
total column density.
More absorption components
could be used to describe the metal lines in this DHAS, 
but the qualitative results will remain
unchanged.  We used the program CLOUDY (Ferland 1993), with an
ionizing background spectrum calculated by Haardt \& Madau (1996), to calculate
the ionization state of the gas in each component.
Table 5 shows the metallicities, temperatures, ionization parameter,
and total hydrogen density in each component calculated from the
CLOUDY simulations.

The equilibrium temperatures calculated from the CLOUDY
simulations can be directly compared to the temperatures determined
from the velocity dispersions of the H, C and Si lines.
In the Blue component, we find a good agreement between the temperatures
shown in Table 4.  But in the Red component the 
CLOUDY equilibrium temperature is smaller than the
temperature determined from the relative line widths, including
$b_{tur}$.
This may indicate another component near the velocity
position of the Red component.
The CLOUDY simulations did not give an acceptable fit to all ions of
Si in the red component.  A single phase of photoionization could not account
for the three column densities observed,
and the calculated [Si/H]$_{red}$ came from Si~II and Si~III alone.
We excluded Si~IV for two reasons: (1) Si~IV(1393) is blended with
a C~IV(1550) at $z=3.1097$, and (2) it is likely that there is another
component of higher ionization that contributes to Si~IV.

We have not attempted to model the metal lines with more than two
components.  Although the column densities of the individual components
would differ, the derived metallicities of the components would not
greatly differ from the results shown in Table 5.  All components would
have a metallicity below 1/100 solar, independent of the model used.

We can now test the TFB assumption that the H and D lines were at
the velocities of the C and Si metal lines.
In general, this assumption adds a systematic error 
to the measurement if the metal lines are not aligned with the
H~I and D~I lines.  
This DHAS is rather unique, 
because
all the metal lines, with different ionizations, show similar
velocity structure.
In Figure 3, the velocity positions of the Blue and Red components 
used by TFB lie
at $\Delta v = 0, 15$ \kms respectively.  In our models with three or
four components, two of the components lie near the velocity positions
of the metal lines, with considerable variation, depending on D/H.  
But in the two component models, the component separation for the most
likely D/H is 
19 \kms, rather then 15 \kms, and the Red hydrogen component
is at a higher velocity position for all values of D/H.
This shows that a systematic offset in the Red component was introduced
in the analysis in TFB by assuming that the H and D lines fell precisely
at the metal line positions in the two component model.
It is difficult to translate this velocity offset
into a D/H change because, as seen in Figure 3, many other
coupled parameters are involved.

\section{PRIMORDIAL D/H}

We have measured the deuterium abundance in the DHAS at $z=3.5722$ towards
Q1937--1009.  Within the statistical uncertainties, 
we conclude that this is the primordial value of D/H produced
by Big Bang Nucleosynthesis (Reeves et al. 1973; \cite{eps76}).
This absorption system is young ($\leq 1\, h^{-1}$ Gyr) and very metal poor 
([X/H] $\leq -2.0$).  These characteristics place strong constraints
on stellar astration of deuterium independent of initial mass functions
and star formation rates (Jedamzik \& Fuller 1997; \cite{fie96}).  
To astrate 
significant amounts of deuterium, high mass stars would
overproduce CNO and Si, and low mass stars would not have time to
complete their evolution.  
With no post-BBN processes to create or destroy
deuterium, our measured value of D/H must represent the primordial value.

\subsection{Other D/H Measurements}

Due to its importance in the standard cosmological model, D/H has
been measured in many astrophysical sites.
In the local interstellar medium, Piskunov et al. (1997) 
find a mean abundance ratio,
$\rm{D/H}_{ISM} = 1.6 \pm 0.2 \times 10^{-5}$, in the lines of sight
toward nearby stars.
Chengalur et al. (1997) have recently detected the D~I 
hyperfine transition in emission 
toward the Galactic anticenter, and find
D/H $=3.9 \pm 1.0 \times 10^{-5}$.
Abundance measurements in meteorites,
the lunar soil, and the atmospheres of the Jovian planets are
consistent with a pre-solar value, 
$\rm{D/H}_{\odot} = 2.6 \pm 1.0 \times 10^{-5}$ (\cite{gau83};
\cite{gei93}; \cite{erc96}).
In the next section, we use the local measurements of D/H with our measurement
at high redshift to discuss the evolution of D/H.

Tytler \& Burles (1997) have made another 
measurement of D/H in a high-redshift Lyman limit system
towards Q1009+2956, at $z=2.504$.
Preliminary 
analysis of this object gives 
D/H = 3.0 $\pm \, 0.6 \times 10^{-5}$, and simulations
of hydrogen contamination suggested the most likely value of D/H
= 2.5 $\pm \, 0.5 \pm 0.4 \times 10^{-5}$ (statistical and systematic error).
The measurement towards Q1009+2956 agrees very well with the result obtained
in this paper towards Q1937--1009. An analysis of Q1009+2956 with
the methods  
presented here will be presented in Burles \& Tytler (1997).

There have been other reports of deuterium detections in high-redshift
QSO absorption systems: towards Q0014+8118 
(\cite{son94}; \cite{car94}; \cite{rug96a}; \cite{rug96b}),
towards Q1202--0725 (\cite{wam96}), towards Q0420--388 (\cite{car96}).
These other systems do not yield measures of
D/H due to either lower quality spectral data, or greater complexities
in the velocity structure of the absorption system.  The next best candidate
system to measure D/H, $z=3.32$ towards Q0014+8118, was shown to have an
interloping hydrogen cloud within 10 \kms of the expected position of deuterium
(Tytler et al. 1996b).
Webb et al. (1997) recently deduced a D/H value at $z=0.701$
towards the low redshift QSO 
1718+4807 using a spectrum obtained with the Hubble Space Telescope
(HST).
Unlike Q1937--1009 and Q1009+2656, only one H~I line was observed, so 
the velocity structure of the H~I is not well known. 
Assuming a single component fit to the \Lya , they find
D/H = $20 \pm 5 \times 10^{-5}$, but this value will remain suggestive
unless confirmed with approved HST observations of the high-order Lyman 
lines.

\subsection{Chemical Evolution of Deuterium}

Many groups have discussed the significance of deuterium 
in galactic chemical evolution, and have used simple models to calculate
the abundance ratio, D/H, as a function of time and metallicity
(\cite{cla85}; \cite{aud74}; \cite{ste92}; \cite{edm94}; \cite{van94}; 
\cite{ste95}; \cite{pra96}; \cite{fie96}; \cite{scu97}).
In the past, deuterium evolution was required
in order to extrapolate the ISM and pre-solar nebula measurements of D/H
back in time, and back to zero metallicity, to constrain the primordial
D/H value.  But now the problem can be inverted, we can use the values
of primordial D/H, pre-solar D/H, and the ISM D/H to constrain 
models of chemical evolution, 
deuterium astration as a function of time and metallicity,
and the fraction of baryons which have been cycled 
through stars in the solar neighborhood.
The ratio of the ISM and primordial values gives
an astration factor in the local ISM,
\begin{equation}
d = {{(\rm{D/H})_{ISM}} \over {(\rm{D/H})_p}} = 0.5 \pm 0.05.
\end{equation}
So we can make the simple statement that one-half of the gas in the 
local ISM has been cycled through stars.  This fraction can
be used to place new limits
on the stellar initial mass function and the amount of inflow
and/or outflow in the local ISM.  When one considers published models
of chemical evolution in the ISM,
the amount of deuterium astration agrees well
with models incorporating standard star formation and some infall
of primordial material (\cite{edm94};
\cite{cla85}; \cite{fie96}).

\subsection{The Baryon Density}

Production of the light elements from Standard Big Bang Nucleosynthesis (SBBN)
depends on a single parameter, $\eta$, the baryon to photon ratio
during SBBN.  Many groups have calculated
the abundance yields of the light elements as a function of $\eta$
(\cite{wag67}; \cite{wal91}; \cite{smi93}; \cite{kra95}; \cite{cop95}).
The abundance yield of deuterium is a single-valued function of $\eta$,
and using output from the Kawano BBN code (\cite{kaw92}), we find
\begin{equation}
\eta = 5.3 \pm 0.3 \pm 0.25 \times 10^{-10},
\end{equation}
where the first error is
the statistical uncertainty in the D/H measurement, and the second error
represents the statistical
uncertainty in the nuclear cross-sections (\cite{smi93}).
It is interesting to note that the magnitude of the 
measurement errors are comparable to the errors in the SBBN calculations.

If we use the present-day photon density determined from the COBE FIRAS
measurements of the Cosmic Microwave Background (CMB, \cite{fix96}),
we can directly calculate the present-day baryon density
\begin{equation}
\Omega_b \, h^2 = 0.0193 \pm 0.0014
\end{equation}
where the error includes both
uncertainties in $\eta$.
This value for the baryon density agrees with recent estimates
of $\Omega_b \, h^2$ from observations of the
CMB power spectrum, the intergalactic medium at
$z \approx 3$, and rich galaxy clusters at $z \approx 0.3$

The power spectrum of anisotropies in the CMB 
determines $\Omega_b$ at the epoch of decoupling, $z \simeq 1000$
(Hu \& White 1996).  The locations and heights of the doppler peaks
in the CMB power spectrum provide constraints on $\Omega_b$. 
The current observational state of the CMB power spectrum does not have
the precision to determine $\Omega_b$, although there are hints of the
first doppler peak and statistical analyses suggest $\Omega_b < 0.28$ 
consistent with SBBN (\cite{lin97}).
The future CMB experiments, including
the satellites, MAP and
Planck, promise to determine $\Omega_b$ to 2\%, which will
provide an independent check on our measurement and a consistency
test for SBBN.

At redshifts $z < 5$, the IGM is ionized, and can be observed through
absorption in high-redshift QSO spectra.  Recent comparisons of
high-resolution spectroscopy of the IGM and large cosmological simulations
require a high baryon density to account for the observed \Lya forest, 
$\Omega_b \geq 0.017 h^{-2}$
(\cite{rau97}; \cite{wei97}; \cite{zha97}).
Other estimates include the baryons in protogalaxies seen in Damped \Lya
absorption with $\Omega_{Damped} \simeq 0.002 h^{-1}$ (\cite{wol95}),
and the baryons in the diffuse IGM which are not seen in \Lya absorption,
$0.0001 \, h^{-1} < \Omega_{diff} < 0.007 \, h^{-3/2}$
(\cite{rei97}).  The inventory of baryons at $z \simeq 3$ 
demands a high cosmological baryon density consistent with our measurements
of D/H.

At low redshift, only a small fraction ($\approx 10$\%) 
of the baryons reside in stars
and visible gas, with estimates ranging from 
$\Omega_* \simeq 0.003 - 0.007 h^{-1}$ (\cite{per92}).
But a fair sample of today's baryons are likely to be in gas heated through
supernova explosions or accretion into the large potential wells of galaxy
clusters.
A recent application of Oort's method to 14 fields containing 
rich galaxy clusters
yields a baryon density in the form of hot gas, 
$\Omega_{gas} \simeq 0.012 - 0.016 h^{-3/2}$ (\cite{car97}),
This result is valid assuming that the baryons and emitted light trace 
the total matter density and that clusters provide a fair sample of the baryon
fraction.  If most of the baryons today are in the form of hot gas, they
must be collisionally ionized (\cite{gia97}), and could have already
been detected in a survey for O~VI at $z \simeq 0.9$ (Burles \& Tytler 1996).

Measurements of the total cosmological matter density $\Omega_m$ imply
that baryons cannot constitute all the matter in the universe.  
A number of methods have been used to estimate 
$\Omega_m$ (c.f. Dekel et al. 1996).
Peculiar velocities of galaxies can be treated as 
cosmic flows induced by large-scale mass distributions.  In particular,
cosmic flows in large-scale voids places a 
 lower limit on the
total mean matter density,
$\Omega_m > 0.3$ (Dekel 1997).
This limit on $\Omega_m$ places an upper limit on the
baryon fraction, $f_b < 0.07$, which agrees with the fraction inferred
from X-ray observations of galaxy clusters (\cite{whi93};
\cite{low96}; \cite{mye97}) and groups (\cite{mul96}).

\subsection{Other Light Elements}

In SBBN, the determination of $\eta$ implies the primordial abundance ratios
for the other light elements.  We now use our D/H measurement to infer
the primordial abundances of the other light elements.
In the following, 
we use the output from the Kawano Code with $\tau = 887$ s, and N$_\nu$ =3.0.
All errors are 67\% confidence and include the uncertainties
in the nuclear cross-sections and neutron lifetime.
Our measurement infers a primordial $^4$He mass fraction Y$_p$ 
\begin{equation}
\rm{Y}_p = 0.247 \pm 0.002.
\end{equation}

The comparison between D and $^4$He provides a crucial test 
of SBBN (\cite{card96}; \cite{hat97}),
but recent measurements of the $^4$He abundance in extragalactic 
metal-poor H~II regions have given discrepant results. 
Izotov et al. (1997) compiled a homogeneous sample of 45 H~II regions
taken mainly from the Second Byurakan Survey (c.f. Izotov et al. 1993),
and found
Y$_p = 0.243 \pm 0.003$ (Izotov et al. 1997), 
which is consistent 
with the value of Y$_p$ we infer.
On the other hand,
Olive et al. (1997) have compiled a large sample of H~II regions
from studies by Olive \& Steigman (1995), Izotov et al. (1994), and Izotov
et al. (1997) and obtain Y$_p = 0.234 \pm 0.002$, 
which is not consistent with Izotov et al. (1997) nor our inferred value. 
The differences in the two $^4$He studies must be properly understood
before a direct comparison can be made between D and $^4$He.
Olive et al. (1997) argue that the
difference can be entirely accounted for in the atomic data sets
and the lowest metallicity region, IZw18.
Skillman \& Kennicutt (1993) observed IZw18 and found 
Y$ = 0.231 \pm 0.005$.
Izotov \& Thuan (1997) reobserved 
IZw18, and they infer Y$ = 0.243 \pm 0.009$.
The different results obtained for this H~II region highlights the
uncertainties in the current observational determinations of Y$_p$.

The primordial value of D/H infers an abundance for $^7$Li,
\begin{equation}
\rm{A(Li)} = 2.5 \pm 0.17,
\end{equation}
where A(Li) = 12 + Log (Li/H).
The abundance of $^7$Li observed in the 
Spite ``plateau" of
warm metal-poor halo stars 
(\cite{spi82}; \cite{spi84}; \cite{reb88}; \cite{tho94}; \cite{bon97})
is lower but consistent with the inferred $^7$Li from D/H.
Bonifacio \& Molaro have recently analyzed a large 
sample of suitable halo stars using infrared measurements to give
a better indication of effective temperatures.
They find 
A(Li) = 2.238 $\pm$ 0.012
$\pm$ 0.05 (statistical and systematic errors, \cite{bon97}), and
no evidence for a dispersion in the plateau abundances,
nor correlations of A(Li) with effective temperature or metallicity.
Our D/H measurements are consistent with the $^7$Li abundance 
found by Bonifacio \& Molaro (1997), but 
large errors in nuclear reaction rates required in the BBN simulations
also allow for non-standard depletion mechanisms in halo stars,
which could have lowered the plateau by as much as 0.6 dex.
(\cite{pis92}; \cite{vau95}).

There do not exist any convincing measurements of a primordial $^3$He
abundance, although great efforts have been made to measure $^3$He
in Galactic H~II regions (\cite{bal94}) and planetary nebulae 
(\cite{bal97}) using the hyperfine
radio line.  Stars can both 
produce or destroy $^3$He, so any attempts to infer the primordial
abundance of $^3$He from galactic measurements are dominated by uncertainties
in the chemical evolution of $^3$He.
If we look at the other side of the problem, 
we can use our inferred primordial value
\begin{equation}
{{^3\rm{He}}\over{\rm{H}}}= 1.2 \pm 0.2 \times 10^{-5}
\end{equation}
to study the chemical evolution of $^3$He.

With the current state of the observations of the light elements 
and the associated uncertainties with inferring primordial abundances,
we conclude that there is no inconsistency between deuterium and
the other light elements with the predictions from SBBN.
Furthermore, the 
determination of $\eta$ with predictions from SBBN should be heavily
weighted towards measurements of D/H in QSO absorption systems
(\cite{ful96}; \cite{sch97}).

\acknowledgments
We are extremely grateful to W.M. Keck foundation which made this
work possible, to Steve Vogt and his team for building HIRES, to
Judy Cohen, Bev Oke, and their team for building LRIS, and 
to Tom Bida, Randy Campbell and Wayne Wack for 
assistance at the telescope.  We are very grateful to Joseph Miller
and the staff of Lick Observatory for the construction, maintenance,
and assistance with the Kast spectrograph, which was made possible by
a generous gift from William and Marina Kast.
We thank 
Christian Cardall, George Fuller, Karsten Jedamzik,
David Kirkman, Martin Lemoine, and Jason X. Prochaska
for many useful conversations.

\clearpage

\section{Figure Captions}

Figure 1: Velocity plots of the Keck HIRES spectrum containing 
Lyman series lines in the D/H absorption system towards Q1937-1009.  
Zero velocity corresponds to a redshift, $z=3.572201$, which is identical
to Figure 1 in TFB.
The Lyman limit region shows Ly-12 through Ly-19 in separate frames to
emphasize the alignment of the entire Lyman series in velocity position.
The data is the solid histogram, each bin corresponds to one pixel.
The gray line tracing the data is the best fit model with 3 components,
Model 4.
The three light tick marks represent the position of the
deuterium Lyman lines, the three solid ticks near $v = 0$ mark the corresponding
hydrogen Lyman lines.  The fourth solid tick marks an additional hydrogen
absorber, which does not show deuterium.
The solid horizontal line at zero flux shows the span of the region of interest
around each Lyman line.  For Ly$\alpha$, the region of interest spans the entire
velocity range shown in Figure 1, while  the other Lyman lines have smaller
regions. 
We use all of the spectrum covering the Lyman series from Ly-12 to the Lyman
limit.
Both the spectrum and model fit have been normalized to the unabsorbed
quasar continuum, which was included in the fitting procedure.  
The 1$\sigma$ error in the data values are shown by the dotted line
near zero flux.

Figure 2a: Spectrum of the \Lya region of the DHAS.  The data and model
fit are displayed as in Fig. 1.
represents the best fit of Model 4.  The dashed line shows the 
is a 5th order Legendre polynomial which gave the best fit to the
unabsorbed QSO continuum.
The tick marks show the positions of all 
absorbers which lie in each region.  The taller tick marks show the 3 main
hydrogen absorbers.  The corresponding deuterium 
lines are located approximately 1 \AA~ blueward of the hydrogen lines, and
are easily distinguishable with the same triple pattern.
The solid line at unity represents our original estimate of the unabsorbed
quasar continuum over the region.  

Figure 2b: Spectrum of the \Lyb region of the DHAS.  The continuum
is a 3rd order Legendre polynomial.  
The gray line is the same model in Fig 2a.

Figure 2c: Same as Fig 2b, spectrum of the \Lyg region of the DHAS.  

Figure 2d: Same as Fig 2b, spectrum of the \Lyd region of the DHAS.  

Figure 2e: Same as Fig 2b, spectrum of the \Lye region of the DHAS.  

Figure 2f: Same as Fig 2b, spectrum of the Ly-6 region of the DHAS.  

Figure 2g: Same as Fig 2b, spectrum of the Ly-7 region of the DHAS.  

Figure 2h: Same as Fig 2b, spectrum of the Ly-9 region of the DHAS.  

Figure 2i: Same as Fig 2b, spectrum of the Ly-Limit region of the DHAS.  

Figure 3a: Results of the fitting procedure for Model 1.  The top panel
shows the \cmin as a function of Log (D/H).  
The remaining panels show 
N, $b$, and $z$ for the main H~I components which gave the best fit.
The parameters are represented as thin lines with different line styles
for each component.  
The bottom panel shows the relative velocity positions of the two components,
$\Delta v = 0$ corresponds to $z=3.572201$.  

Figure 3b: Same as Fig 3a, but for Model 2 with three components.

Figure 3c: Same as Fig 3a, but for Model 3 with two components.

Figure 3d: Same as Fig 3a, but for Model 4 with three components.

Figure 3e: Same as Fig 3a, but for Model 5 with four components.

Figure 3f: Same as Fig 3a, but for Model 6 with three components and
contamination at D-\Lya.

Figure 3g: Same as Fig 3a, but for Model 7 with three components and
contamination at D-\Lyb.

Figure 4: 95\% confidence regions of D/H for the seven models in Table 3.
The central vertical tick marks show the values of D/H which gave the best
fit in each model.

Figure 5: Same as Fig 2b, but with the best fit for Model 7. 



\begin{deluxetable}{ccccc}
\tablecolumns{5}
\tablenum{1}
\tablewidth{0pc}
\tablecaption{Spectral Regions used in D/H Measurement }
\tablehead{
\colhead{Region} & \colhead{$\lambda_{min}$} & \colhead{$\lambda_{max}$} & 
\colhead{Pixels} & \colhead{Order\tablenotemark{a}}}
\startdata
\Lya & 5546.08 & 5568.57 & 301 & 5 \cr
\Lyb & 4687.50 & 4692.00 & 70 & 3 \cr
\Lyg & 4445.60 & 4451.56 & 98 & 3 \cr
\Lyd & 4340.40 & 4346.00 & 94 & 3 \cr
\Lye & 4284.70 & 4293.10 & 145 & 3 \cr
Ly-6  & 4253.30 & 4257.06 & 63  & 3 \cr
Ly-7  & 4232.15 & 4239.10 & 120 & 3 \cr
Ly-9  & 4208.35 & 4213.10 & 82 & 3 \cr
Ly-Limit & 4178.30 & 4196.56 & 325 & 3 \cr
\enddata
\tablenotetext{a}{Order of Legendre polynomial used for the continuum}
\end{deluxetable}

\clearpage

\begin{deluxetable}{ccc|ccc|ccc}
\tablecolumns{9}
\tablenum{2}
\tablewidth{0pc}


\tablecaption{Group 1 Lines in D/H Models}
\tablehead{
\colhead{ Log N} & \colhead{$b$} & \colhead{$z$} &
\colhead{ Log N} & \colhead{$b$} & \colhead{$z$} &
\colhead{ Log N} & \colhead{$b$} & \colhead{$z$} }
\startdata
12.73 & 30.3 & 3.58001 & 14.27 & 40.8 & 2.65975 & 14.34 & 33.1 & 2.48516 \cr
12.59 & 32.0 & 3.57881 & 14.76 & 31.8 & 2.65795 & 14.16 & 37.5 & 2.48387 \cr
12.88 & 18.2 & 3.57864 & 13.44 & 16.7 & 2.57458 & 12.64 & 19.8 & 2.48218 \cr
13.54 & 21.0 & 3.57846 & 14.59 & 35.4 & 2.57384 & 13.01 & 19.3 & 2.46537 \cr
13.39 & 17.1 & 3.57781 & 14.08 & 62.5 & 2.57216 & 12.51 & 11.0 & 2.46489 \cr
13.12 & 27.2 & 3.57749 & 13.84 & 26.4 & 2.57127 & 14.57 & 47.3 & 2.46359 \cr
13.62 & 20.7 & 3.57673 & 13.58 & 32.3 & 2.53110 & 12.70 & 18.4 & 2.46265 \cr
12.28 & 22.8 & 3.57584 & 13.43 & 24.0 & 2.53043 & 12.93 & 19.1 & 2.46216 \cr
13.24 & 63.4 & 3.57503 & 13.70 & 62.5 & 2.52962 & 13.84 & 70.2 & 2.45247 \cr
12.92 & 20.7 & 3.57381 & 12.31 &  9.3 & 2.52853 & 13.75 & 38.2 & 2.45109 \cr
15.40 & 31.1 & 3.57295 & 13.62 & 22.1 & 2.52785 & 14.17 & 34.1 & 2.44969 \cr
13.20 & 44.0 & 3.56964 & 13.39 & 43.7 & 2.52595 & 12.52 &  5.9 & 2.44885 \cr
12.39 & 22.7 & 3.56837 & 13.59 & 33.7 & 2.52506 & 14.00 & 28.7 & 2.44785 \cr
12.43 & 16.3 & 3.56760 & 14.04 & 28.6 & 2.50091 & 13.86 & 22.0 & 2.44761 \cr
13.63 & 34.5 & 3.56699 & 13.60 & 35.5 & 2.50053 & 13.13 &  9.3 & 2.44647 \cr
14.00 & 26.8 & 3.56640 & 12.80 & 10.7 & 2.49969 & 14.06 & 31.7 & 2.44450 \cr
14.31 & 32.9 & 3.56495 & 13.50 & 21.1 & 2.49948 & 13.53 & 20.9 & 2.44366 \cr
13.05 & 20.6 & 3.56427 & 13.18 & 30.9 & 2.49919 & 13.11 & 22.2 & 2.44332 \cr
13.42 & 33.3 & 3.56333 & 13.24 & 37.0 & 2.48705 & 14.00 & 26.4 & 2.44115 \cr
12.30 &  6.8 & 2.66147 & 13.18 & 21.4 & 2.48646 & 12.96 &  7.5 & 2.44042 \cr
12.96 & 20.5 & 2.66101 & 13.32 & 38.8 & 2.48615 & 13.60 & 28.5 & 2.43971 \cr
&  & & & & & 13.83 &  8.9 & 2.43868 \cr
\enddata
\end{deluxetable}

\clearpage


\begin{deluxetable}{cccccccc}
\tablecolumns{5}
\tablenum{3}
\tablewidth{0pc}
\tablecaption{D/H Absorption Models}
\tablehead{
\colhead{Model} & \colhead{Components\tablenotemark{a}}  & \colhead{D/H ($-2\sigma$)\tablenotemark{b}} &
\colhead{D/H($\chi^2_{min}$)} & \colhead{D/H ($+2\sigma$)\tablenotemark{b}} & \colhead{$\chi^2_{min}$} &
\colhead{$\nu$\tablenotemark{c}}}
\startdata
 1  & 2 & $-4.55$ & $-4.51$ & $-4.46$ & 424.9 & 1097  \cr
 2  & 3 & $-4.55$ & $-4.48$ & $-4.41$ & 420.6 & 1093  \cr
 3  & 2 & $-4.55$ & $-4.51$ & $-4.47$ & 394.3 & 1069  \cr
 4  & 3 & $-4.56$ & $-4.49$ & $-4.44$ & 387.2 & 1065  \cr
 5  & 4 & $-4.55$ & $-4.48$ & $-4.42$ & 383.5 & 1061  \cr
 6  & 3\tablenotemark{d} & $-4.87$ & $-4.59$ & $-4.43$ & 382.0 & 1062  \cr
 7  & 3\tablenotemark{e} & $-4.57$ & $-4.49$ & $-4.42$ & 379.1 & 1062  \cr
\enddata
\tablenotetext{a}{Number of main components in fit}
\tablenotetext{b}{95\% confidence levels from $\chi^2$ test}
\tablenotetext{c}{degrees of freedom with 1298 pixels}
\tablenotetext{d}{Additional contaminating hydrogen component at \Lya}
\tablenotetext{e}{Additional contaminating hydrogen component at \Lyb}
\end{deluxetable}



\begin{deluxetable}{ccc}
\tablecolumns{3}
\tablenum{4}
\tablewidth{0pc}
\tablecaption{Column Densities of Metals}
\tablehead{
\colhead{ } & \colhead{Blue Component} & \colhead{Red Component}}
\startdata
C I & \multispan{2}{\hfil $<12.4 \; (2\, \sigma)$ \hfil} \cr
C II & 12.70 $\pm$ 0.08 & 13.27 $\pm$ 0.03 \cr
C III & 13.36 $\pm$ 0.09 & 13.82 $\pm$ 0.19 \cr
C IV & 12.22 $\pm$ 0.16 & 12.61 $\pm$ 0.11 \cr
\noalign{\vskip 15pt}

N I & \multispan{2}{\hfil $<12.6 \; (2\, \sigma)$ \hfil} \cr
N II& \multispan{2}{\hfil $<13.9 \; (2\, \sigma)$ \hfil} \cr
N III\tablenotemark{a} & 13.15 $\pm$ 0.71 & 13.84 $\pm$ 0.39  \cr
N V & \multispan{2}{\hfil $<12.4 \; (2\, \sigma)$ \hfil} \cr
\noalign{\vskip 15pt}

O I & \multispan{2}{\hfil $<12.6 \; (2\, \sigma)$ \hfil} \cr
\noalign{\vskip 15pt}

Si II & 11.76 $\pm$ 0.07 & 12.41 $\pm$ 0.02  \cr
Si III & 12.73 $\pm$ 0.20 & 13.20 $\pm$ 0.05  \cr
Si IV\tablenotemark{b} & 12.12 $\pm$ 0.12 & 13.01 $\pm$ 0.02  \cr
\noalign{\vskip 15pt}

Fe II & 11.58 $\pm$ 0.44 & 12.41 $\pm$ 0.10  \cr
Fe III & 12.83 $\pm$ 0.58 & 13.30 $\pm$ 0.10  \cr
\enddata
\tablenotetext{a}{Absorption feature is blended in \Lya forest}
\tablenotetext{b}{Si~IV(1393) is blended with C~IV(1550) at $z=3.1097$}
\end{deluxetable}


\begin{deluxetable}{ccc}
\tablecolumns{3}
\tablenum{5}
\tablewidth{0pc}

\tablecaption{Metallicity and Ionization State of DHAS}
\tablehead{
\colhead{ } & \colhead{Blue Component} & \colhead{Red Component}}
\startdata
 [C/H]     & $-3.0$ & $-2.1$ \cr
 [N/H]     & $\leq -2.7$ & $\leq -2.0$ \cr
 [O/H] & $<-0.9$ & $<-0.9$ \cr
 [Si/H]     & $-2.7$ & $-1.9$ \cr
 [Fe/H] & $<-1.5$ & $<-0.6$ \cr
 Log U & $-2.9$ & $-3.0$ \cr
 Log H~I/H    & $-2.35$ & $-2.29$ \cr
 Log $n_{\rm{H}}$\tablenotemark{a} (cm$^{-3}$) & $-1.70$ & $-1.60$ \cr
 L(kpc)   & 1.5 & 0.6 \cr
 T$_{b}$\tablenotemark{b} & 1.62 $\pm$ 0.09 & 2.36 $\pm$ 0.09 \cr
 T$_{e}$\tablenotemark{c} & 1.72 & 1.67 \cr
 b$_{tur}$ & 4.8 $\pm$ 0.8 & 8.4 $\pm$ 0.4 \cr
\enddata
\tablenotetext{a}{Corresponding to Log $J_0 = -21.3$}
\tablenotetext{b}{Determined from component line widths}
\tablenotetext{c}{Photoionization equilibrium temperature}
\end{deluxetable}

\begin{figure}
\figurenum{1}
\centerline{
\psfig{figure=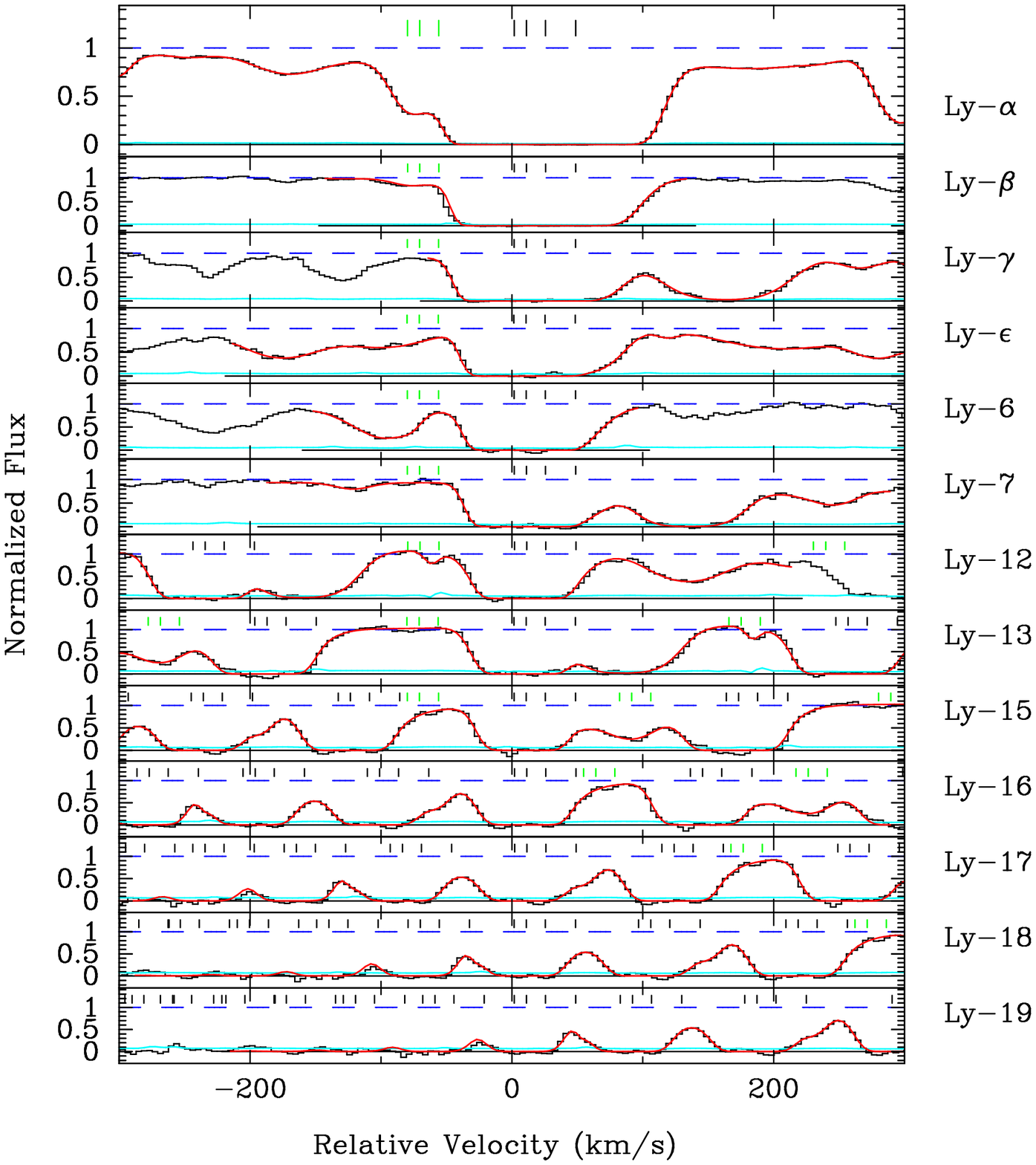,height=\textheight}}
\end{figure}

\begin{figure}
\figurenum{2a}
\centerline{
\psfig{figure=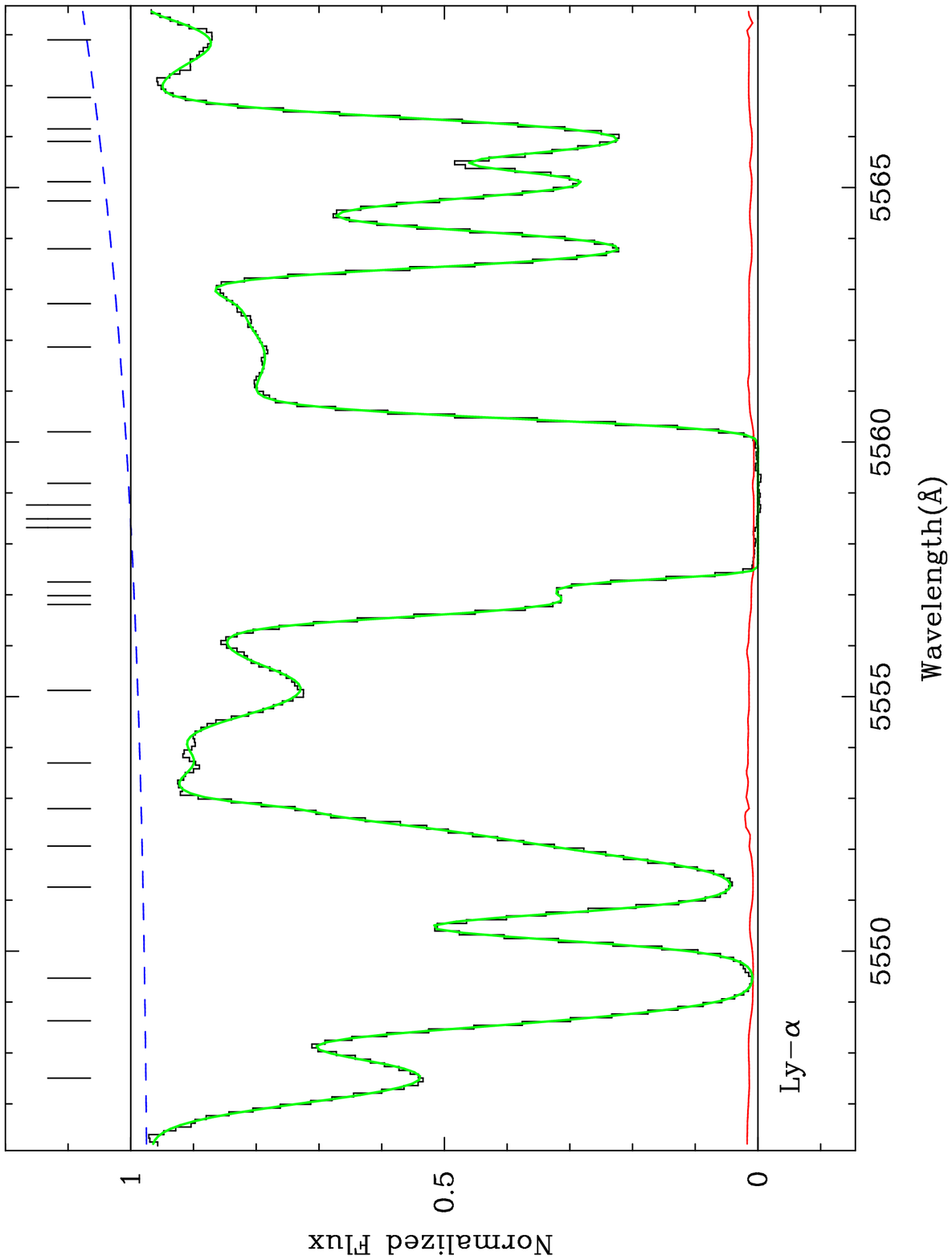,height=\textheight}}
\end{figure}

\begin{figure}
\figurenum{2b}
\centerline{
\psfig{figure=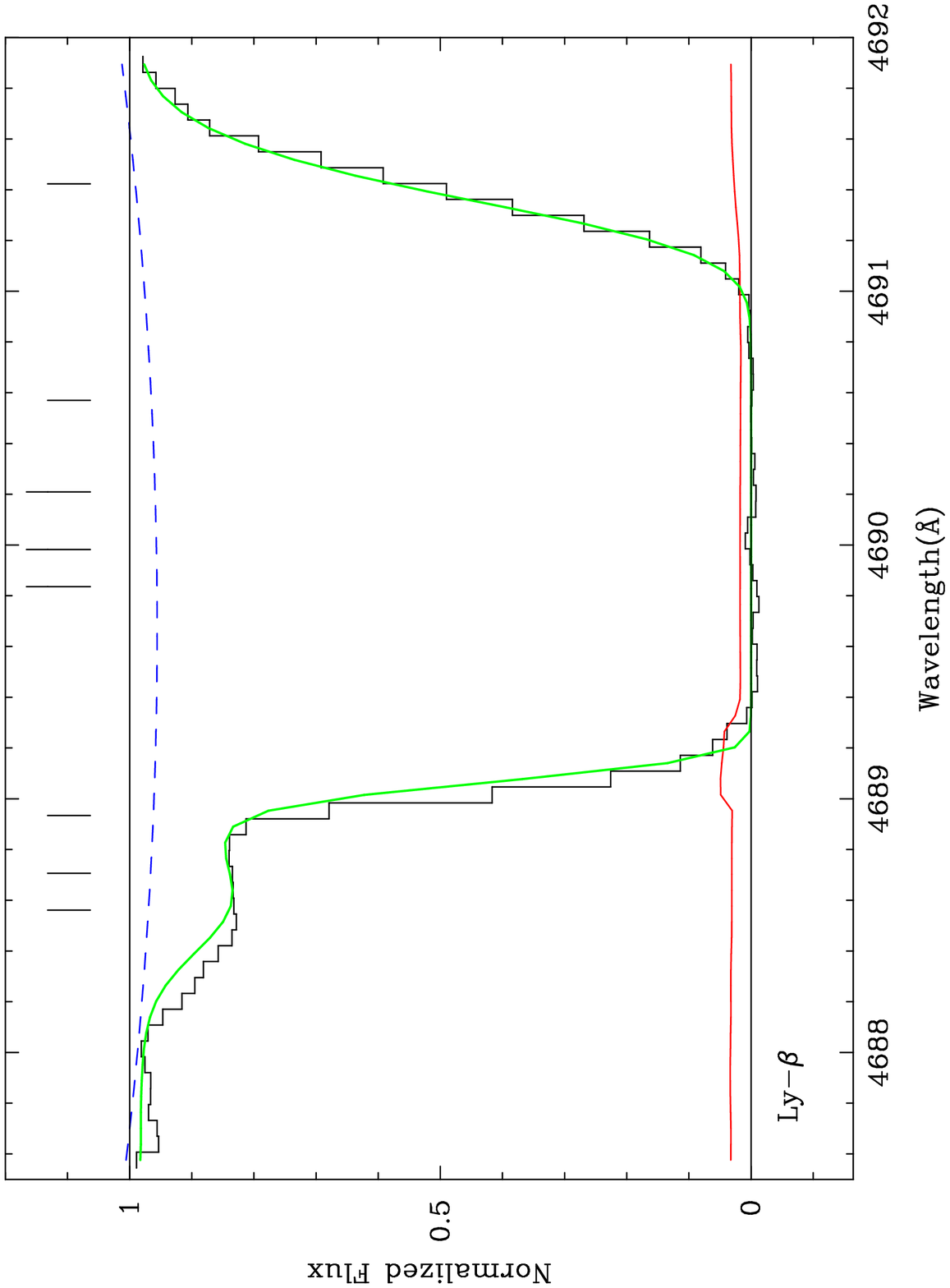,height=\textheight}}
\end{figure}

\begin{figure}
\figurenum{2c}
\centerline{
\psfig{figure=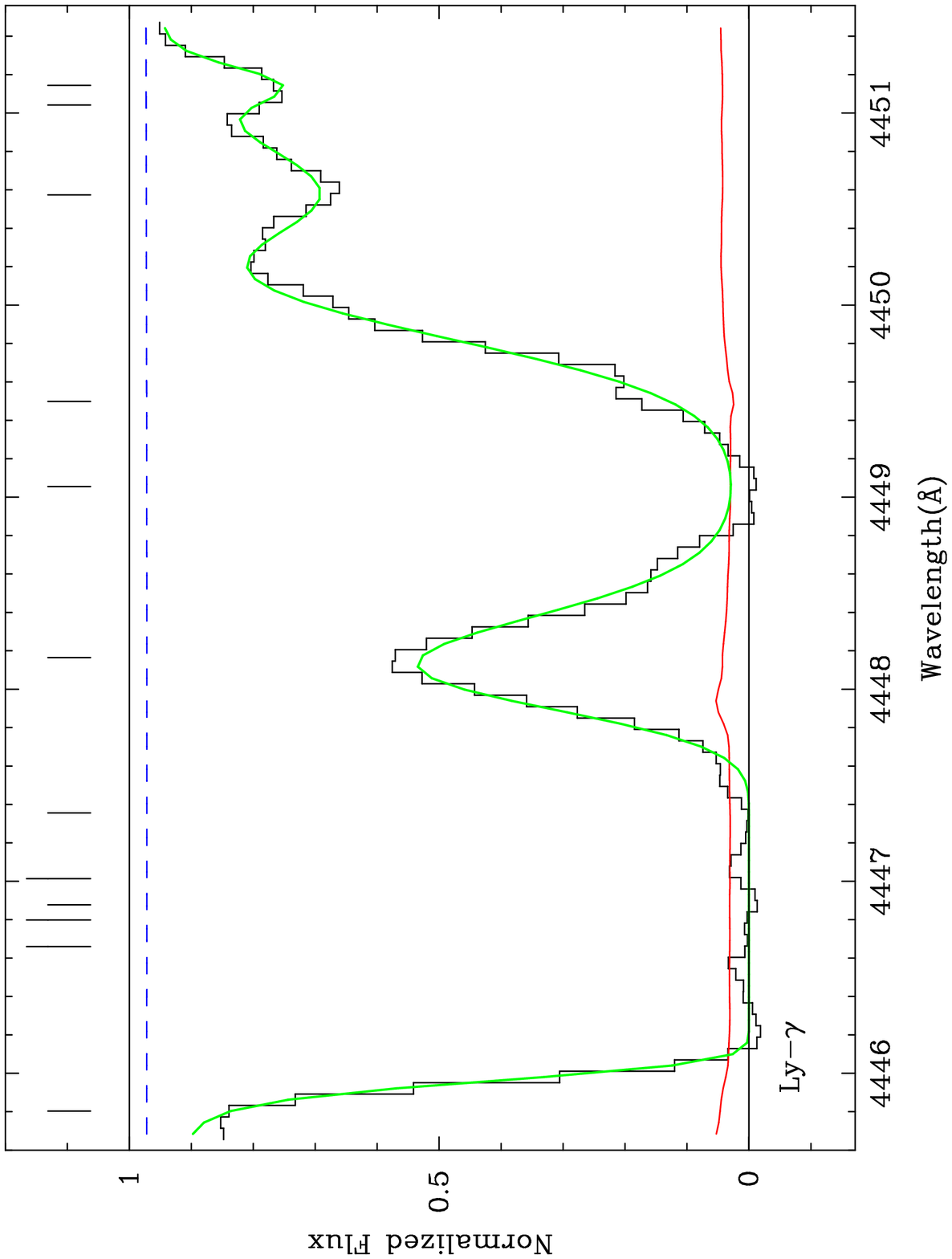,height=\textheight}}
\end{figure}

\begin{figure}
\figurenum{2d}
\centerline{
\psfig{figure=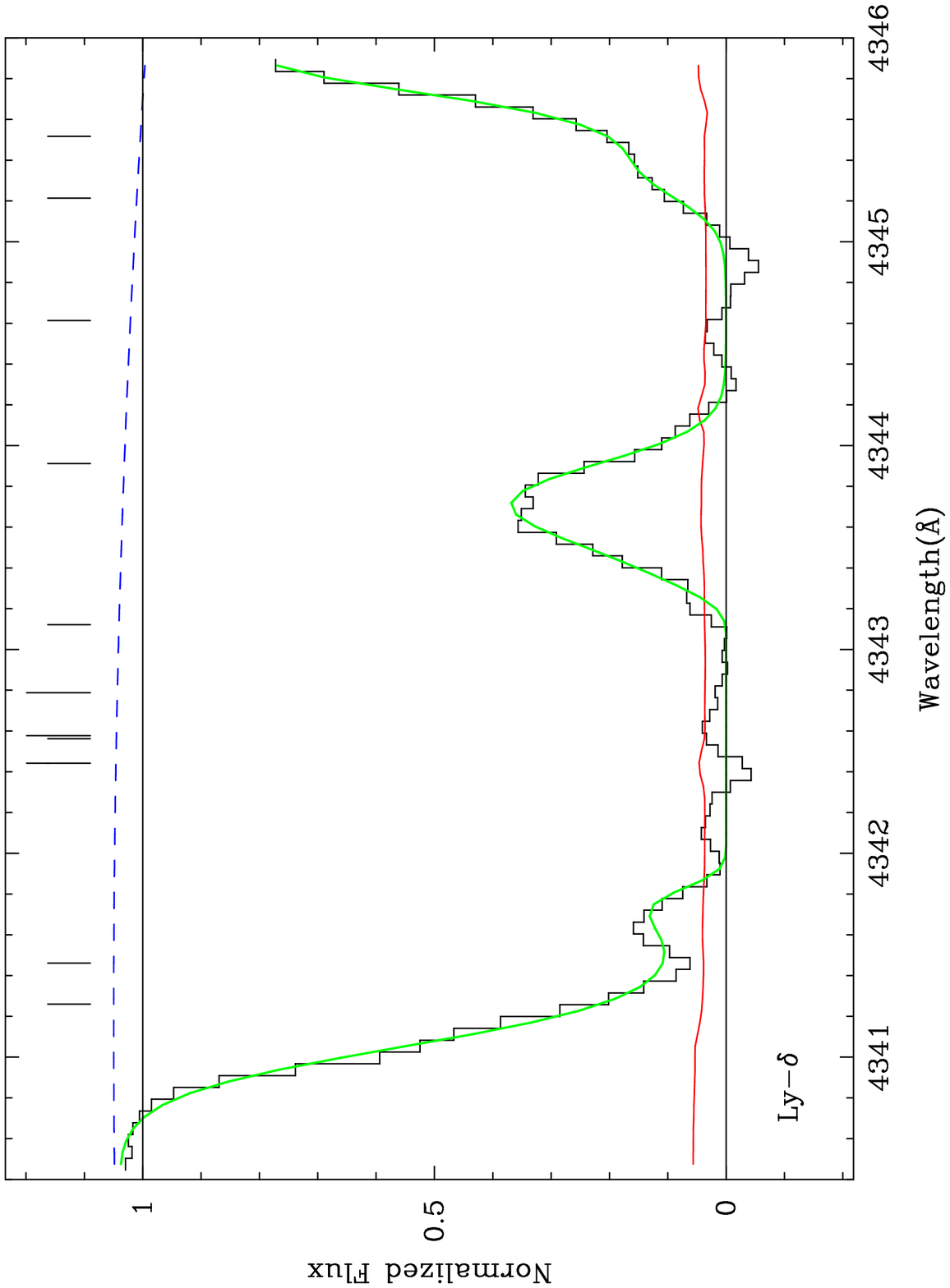,height=\textheight}}
\end{figure}

\begin{figure}
\figurenum{2e}
\centerline{
\psfig{figure=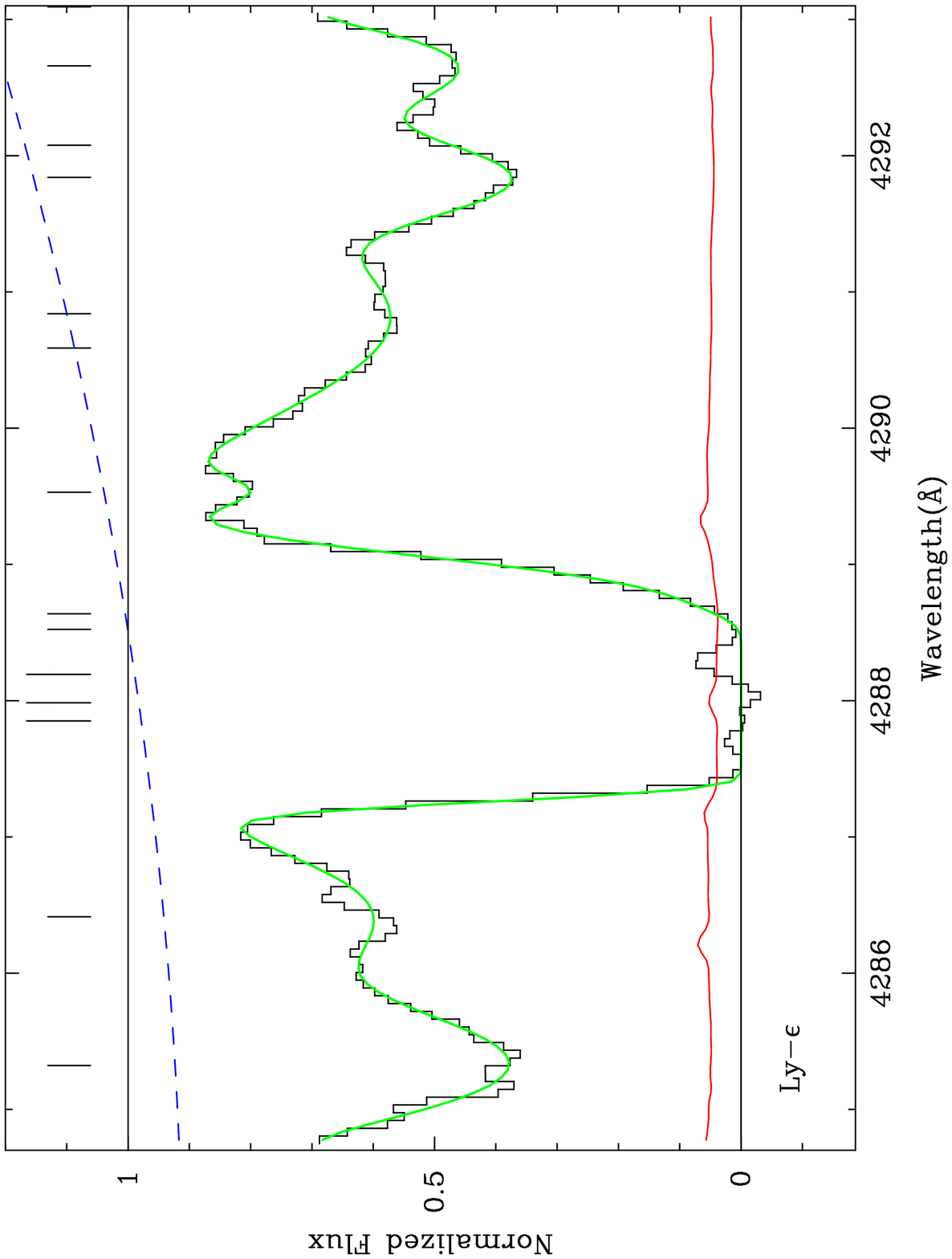,height=\textheight}}
\end{figure}

\begin{figure}
\figurenum{2f}
\centerline{
\psfig{figure=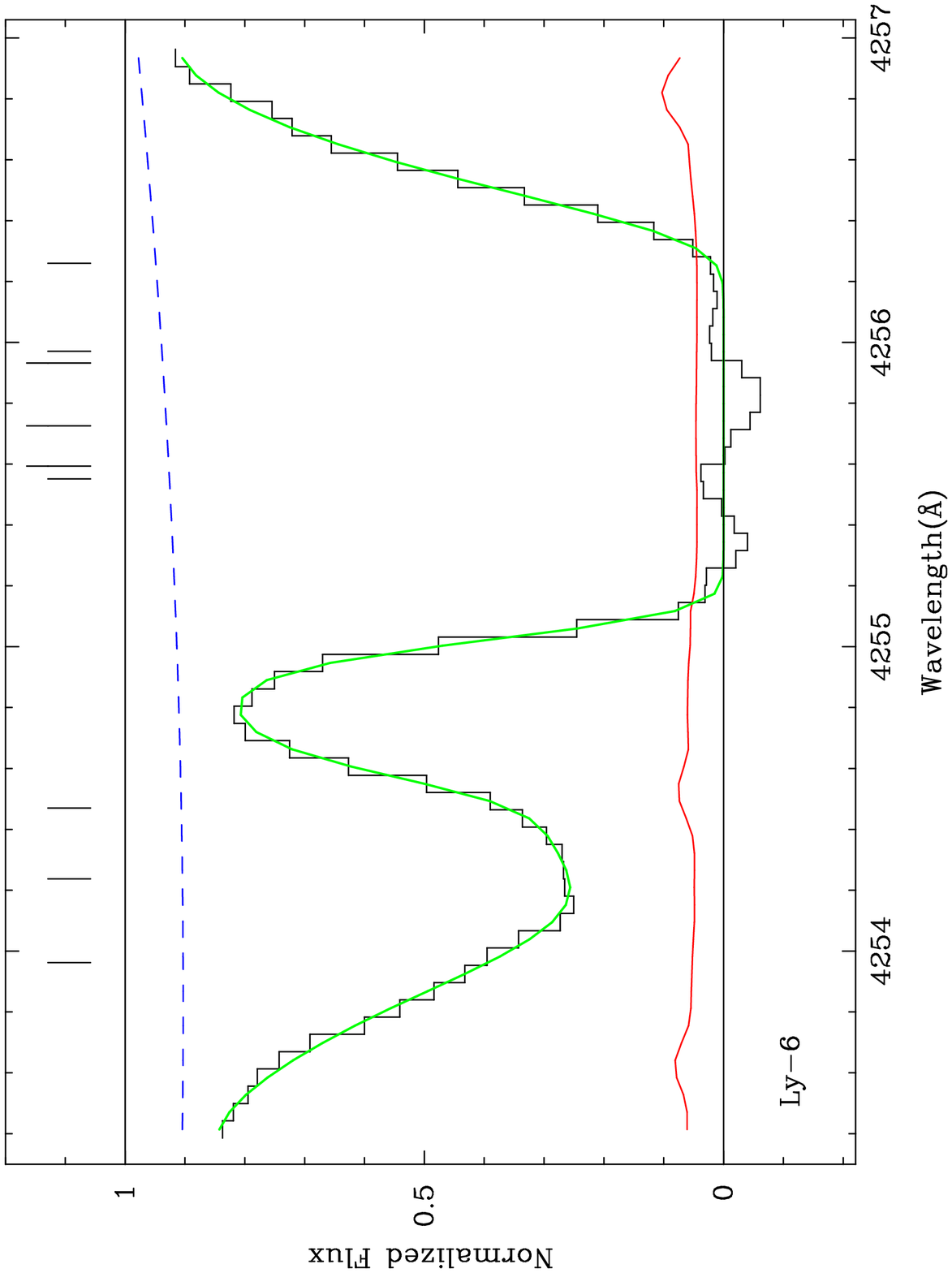,height=\textheight}}
\end{figure}

\begin{figure}
\figurenum{2g}
\centerline{
\psfig{figure=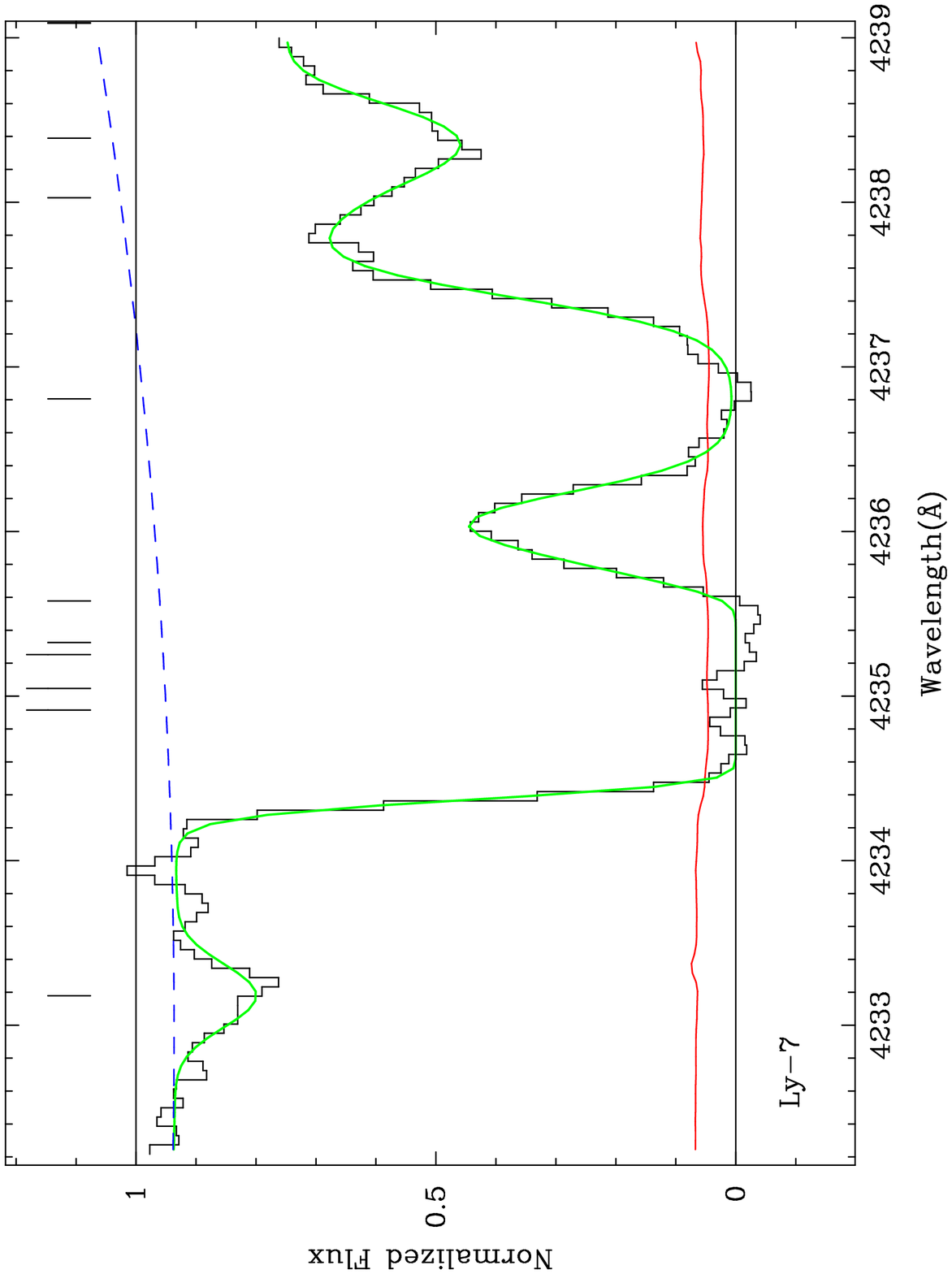,height=\textheight}}
\end{figure}

\begin{figure}
\figurenum{2h}
\centerline{
\psfig{figure=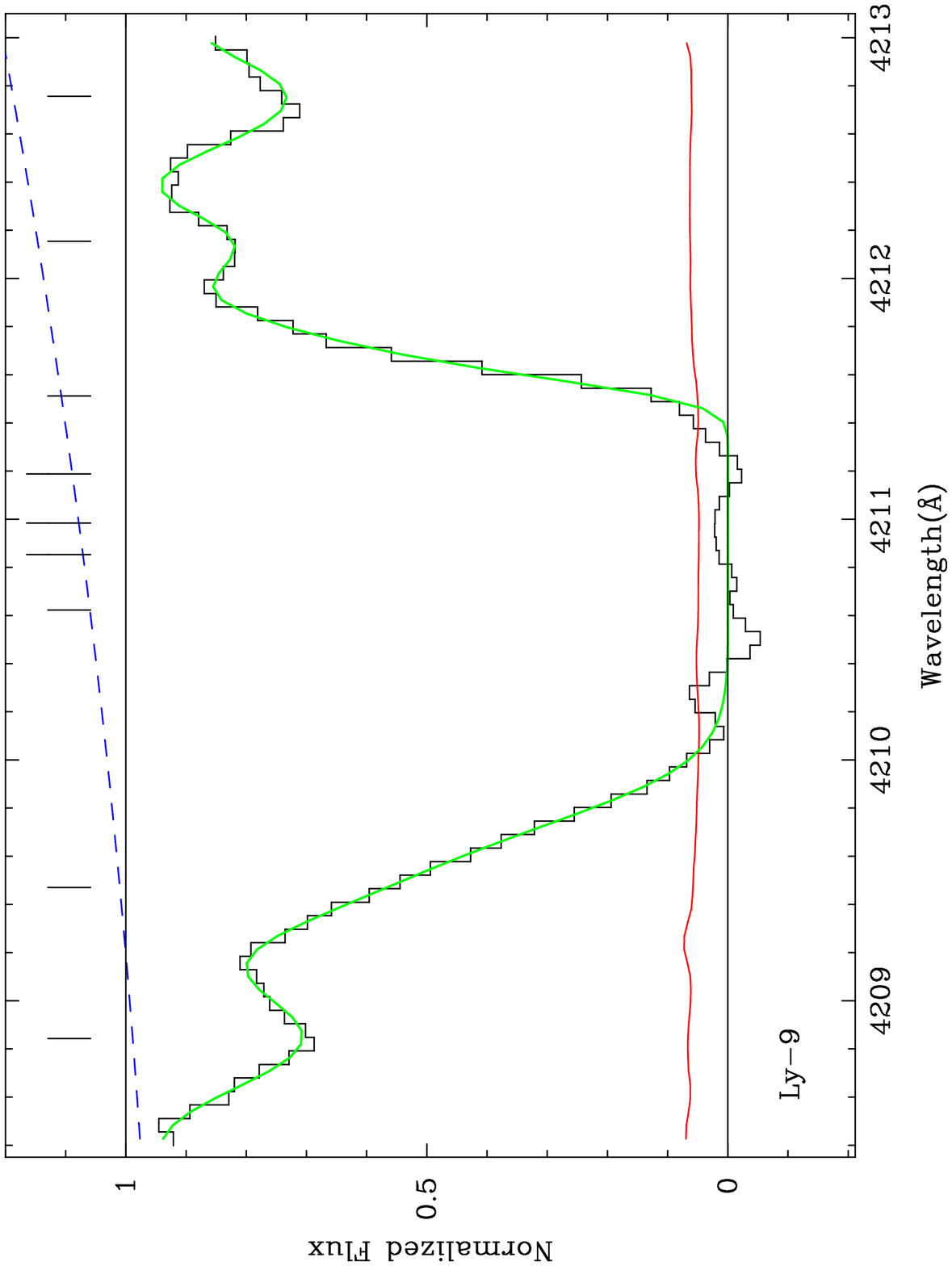,height=\textheight}}
\end{figure}

\begin{figure}
\figurenum{2i}
\centerline{
\psfig{figure=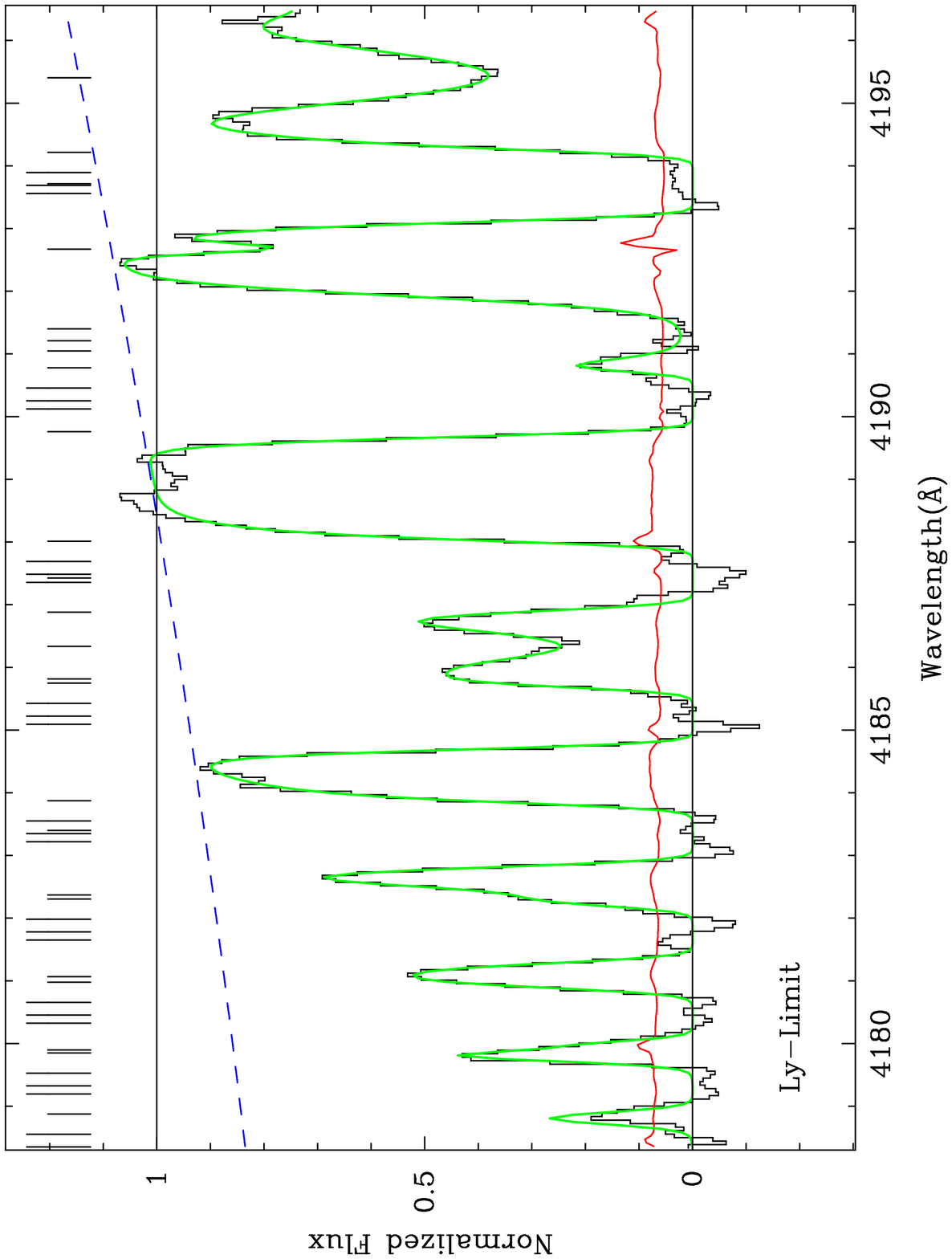,height=\textheight}}
\end{figure}

\clearpage

\begin{figure}
\figurenum{3a}
\centerline{
\psfig{figure=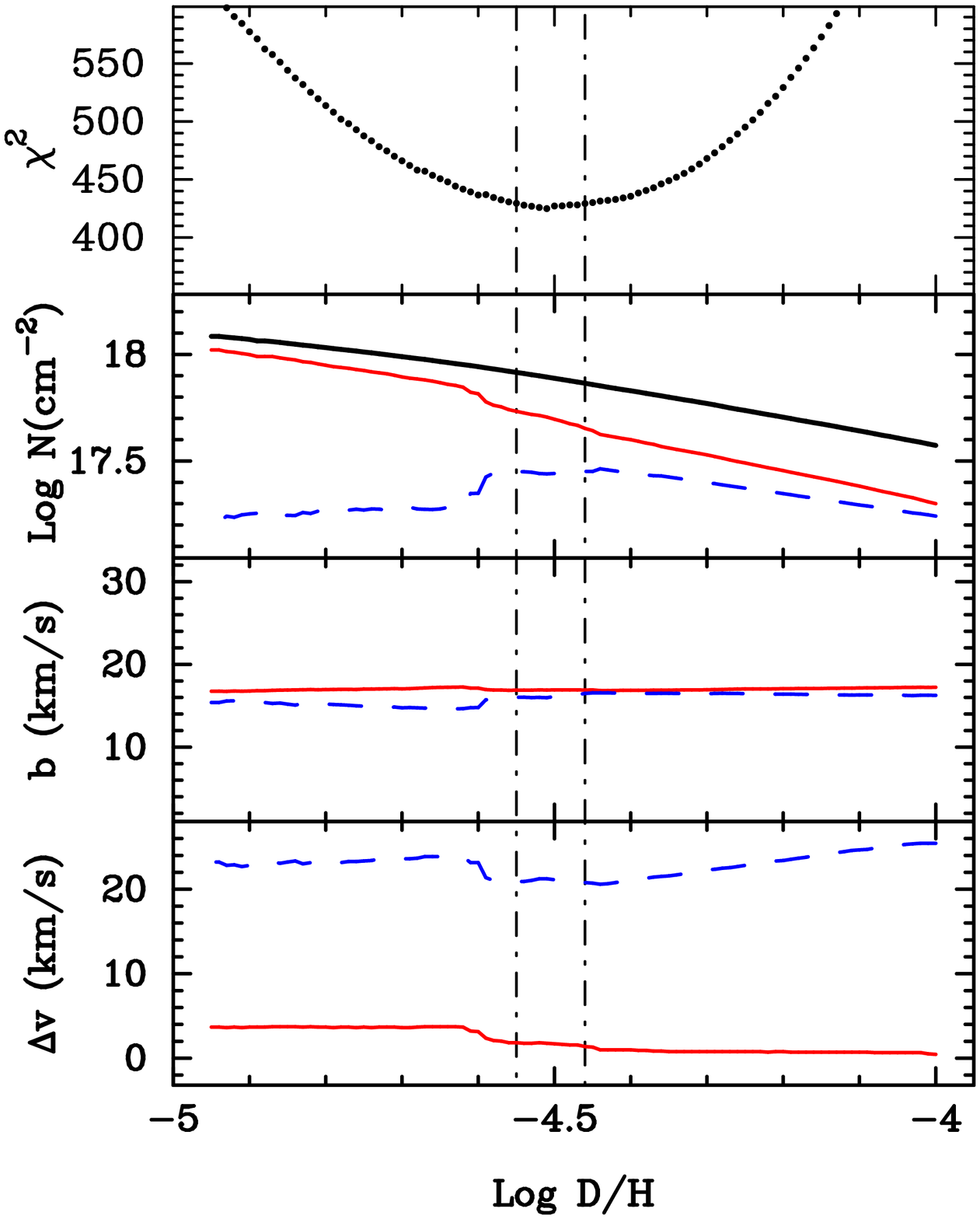,height=\textheight}}
\end{figure}

\begin{figure}
\figurenum{3b}
\centerline{
\psfig{figure=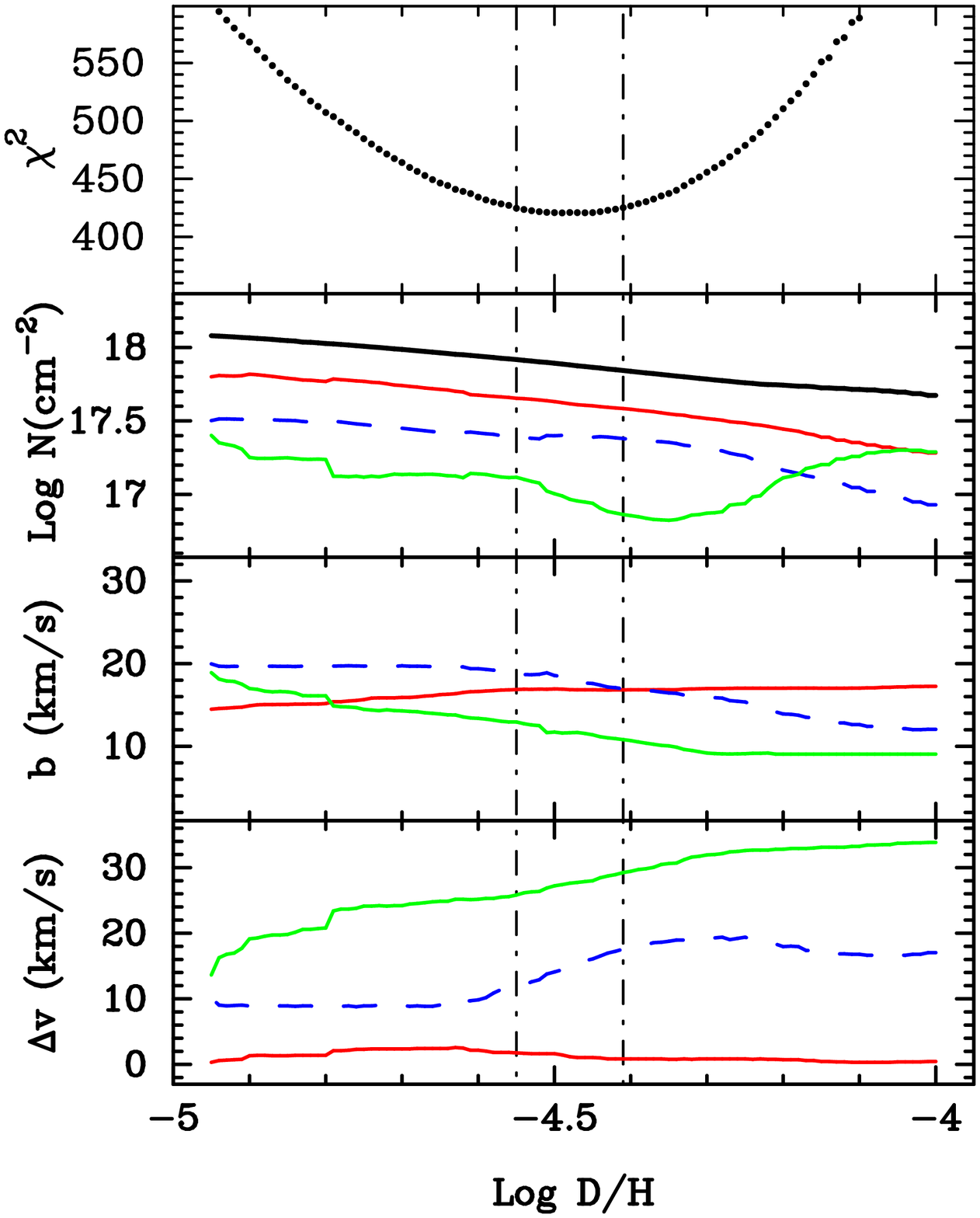,height=\textheight}}
\end{figure}

\begin{figure}
\figurenum{3c}
\centerline{
\psfig{figure=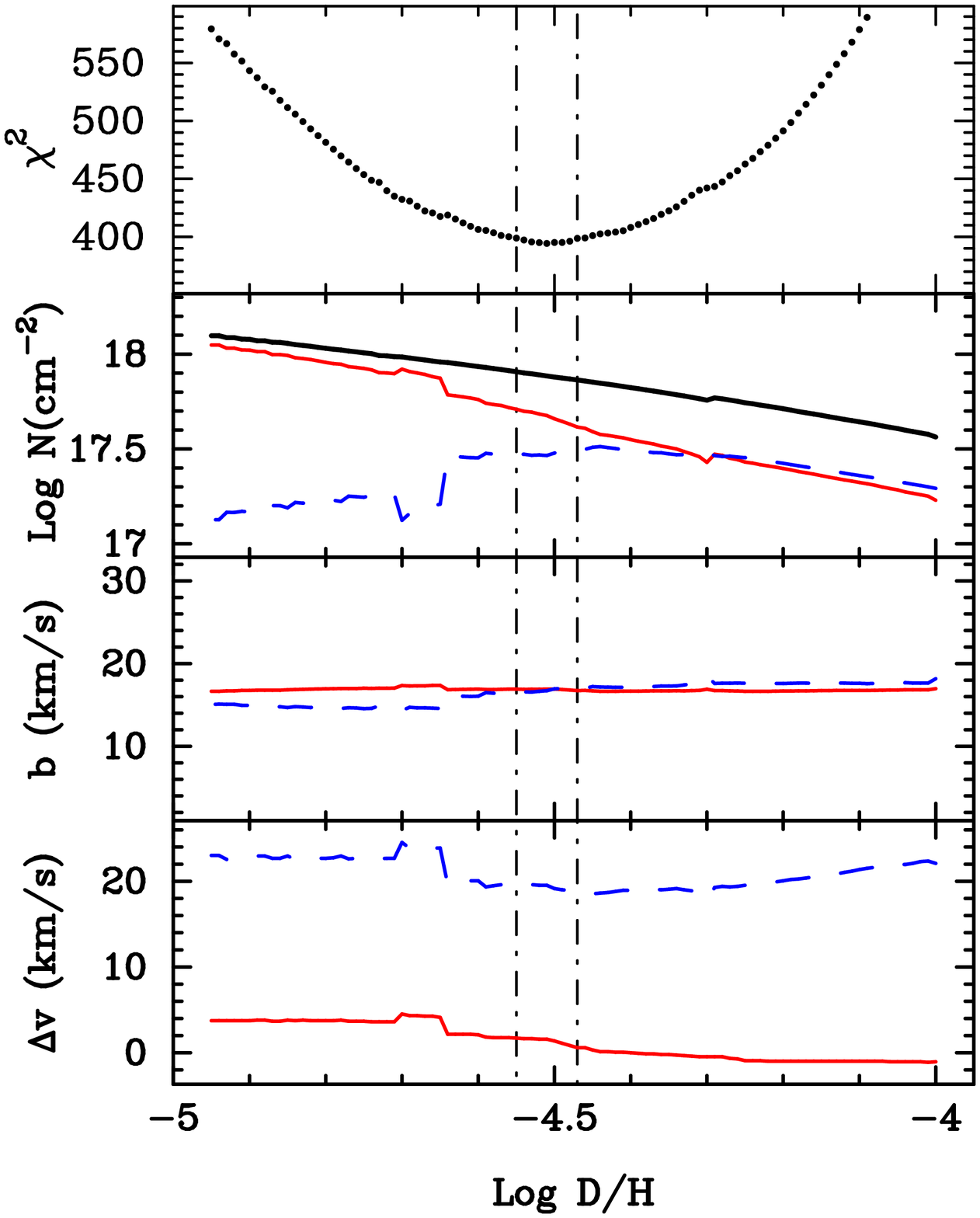,height=\textheight}}
\end{figure}

\begin{figure}
\figurenum{3d}
\centerline{
\psfig{figure=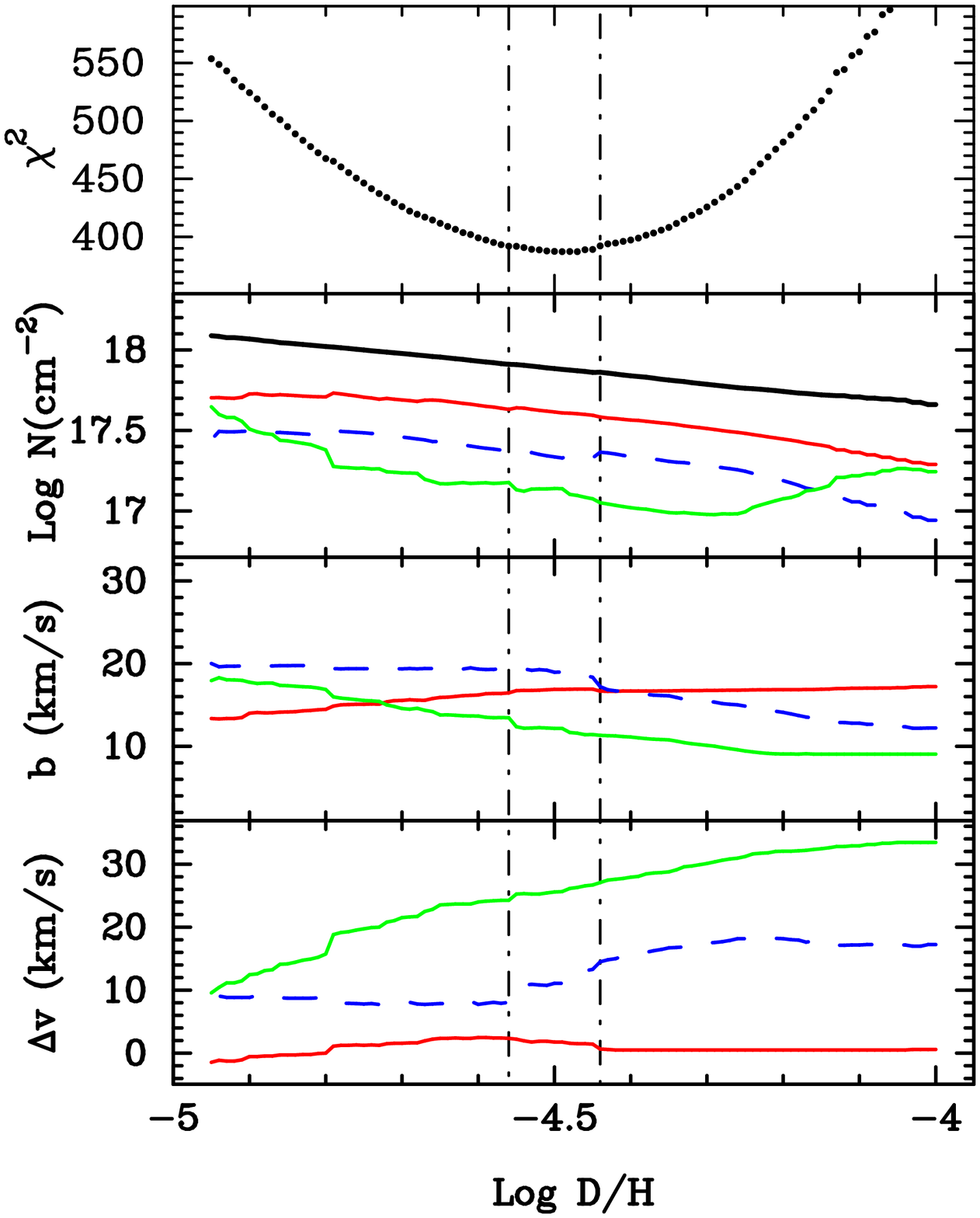,height=\textheight}}
\end{figure}

\begin{figure}
\figurenum{3e}
\centerline{
\psfig{figure=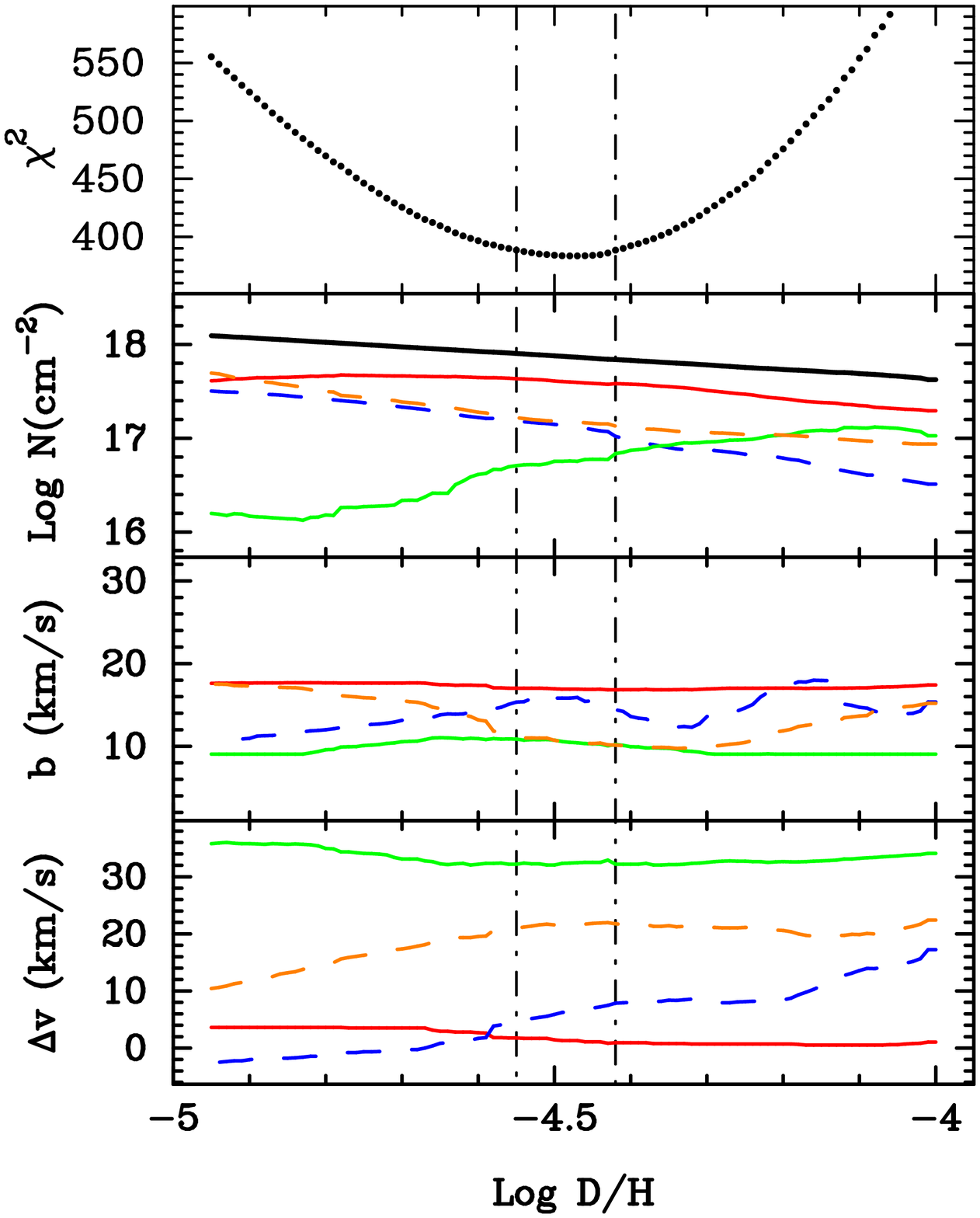,height=\textheight}}
\end{figure}

\begin{figure}
\figurenum{3f}
\centerline{
\psfig{figure=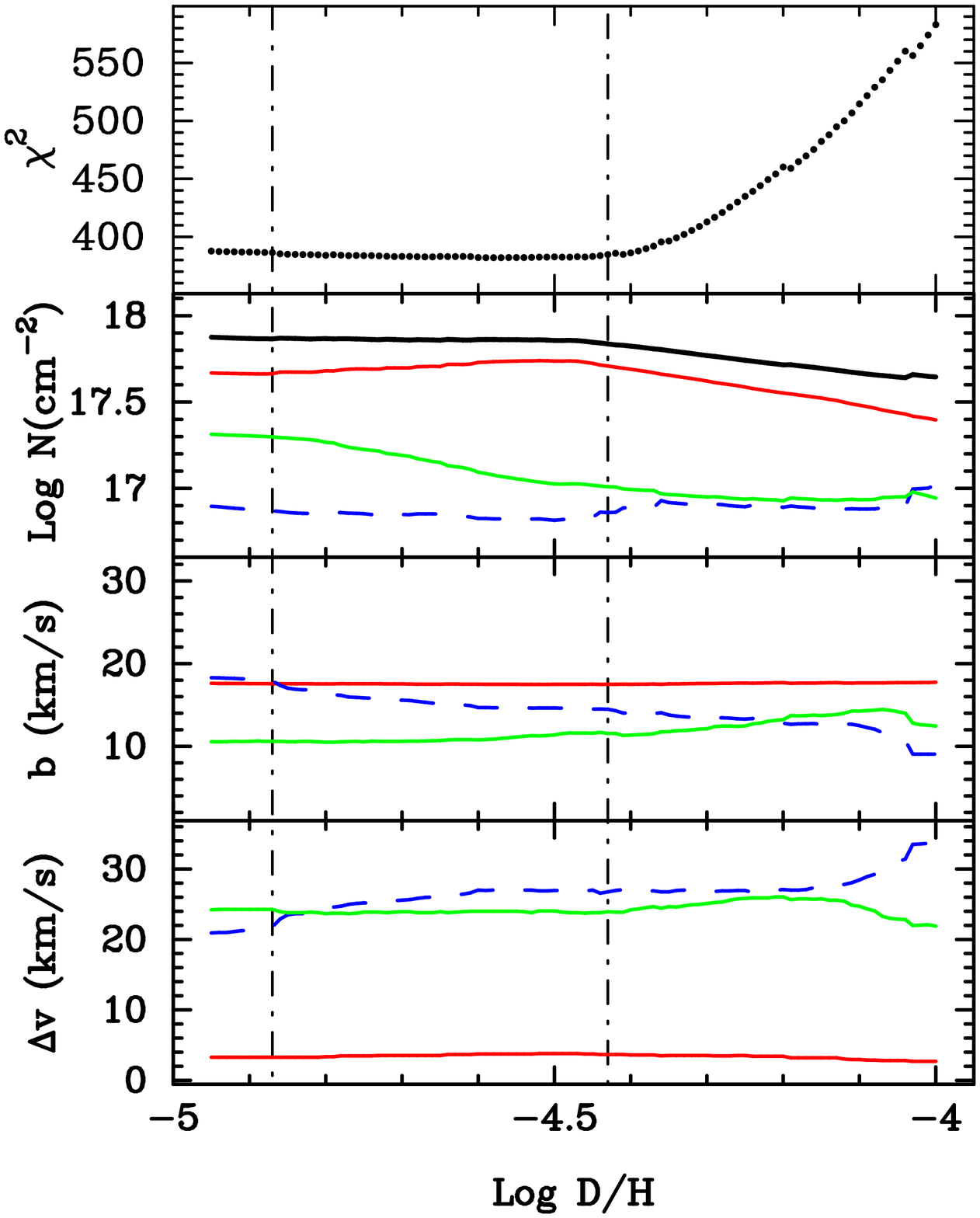,height=\textheight}}
\end{figure}

\begin{figure}
\figurenum{3g}
\centerline{
\psfig{figure=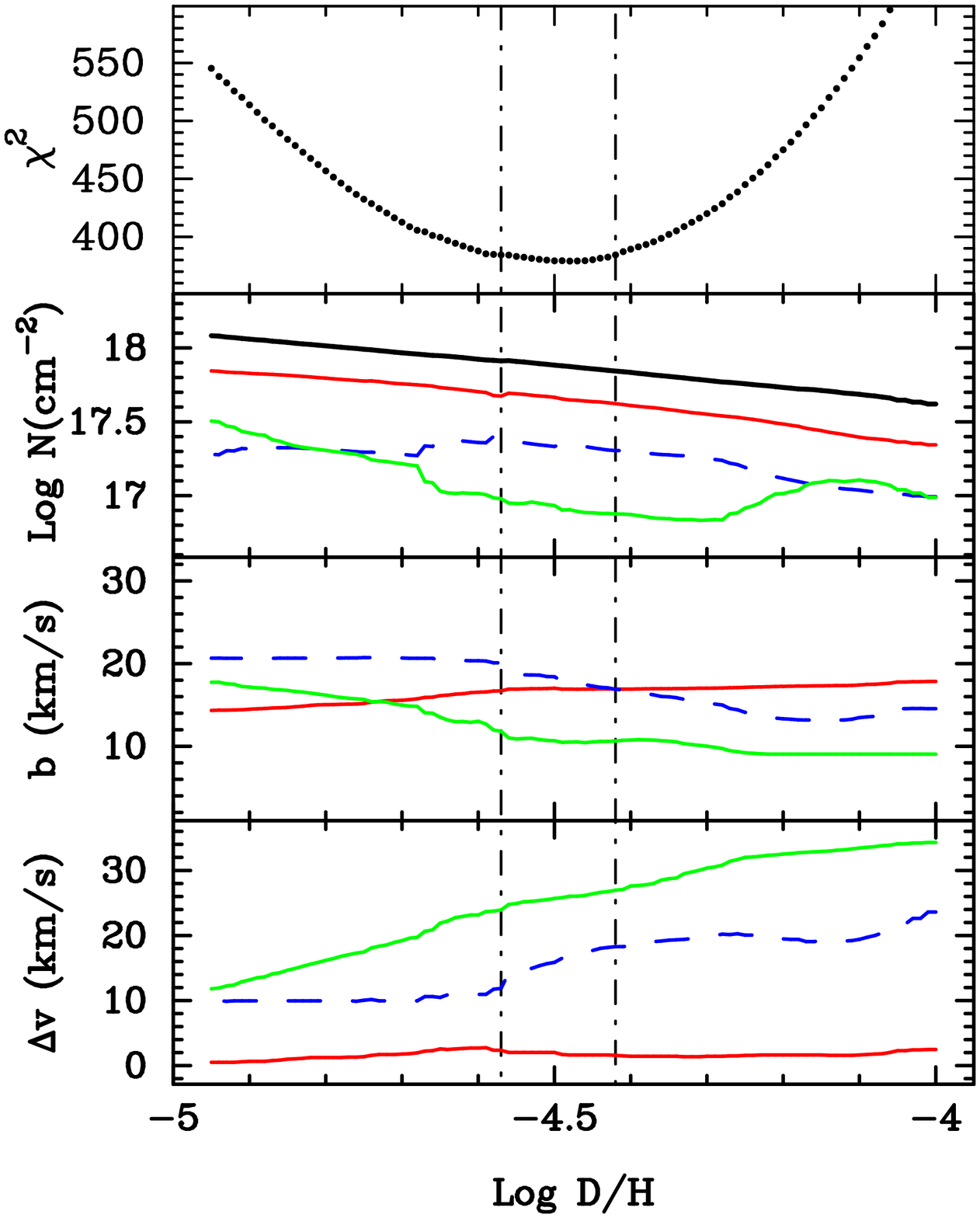,height=\textheight}}
\end{figure}

\clearpage

\begin{figure}
\figurenum{4}
\centerline{
\psfig{figure=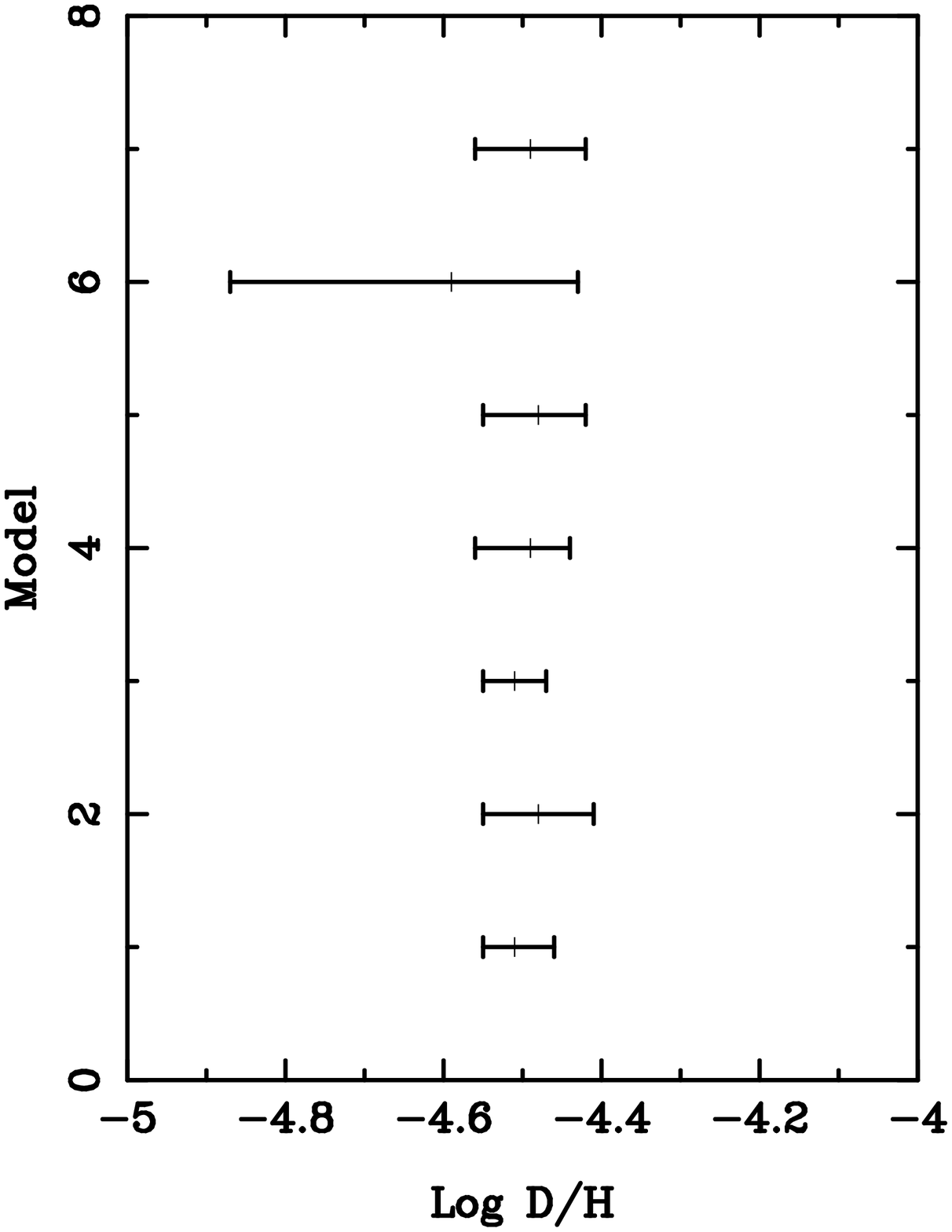,height=\textheight}}
\end{figure}

\begin{figure}
\figurenum{5}
\centerline{
\psfig{figure=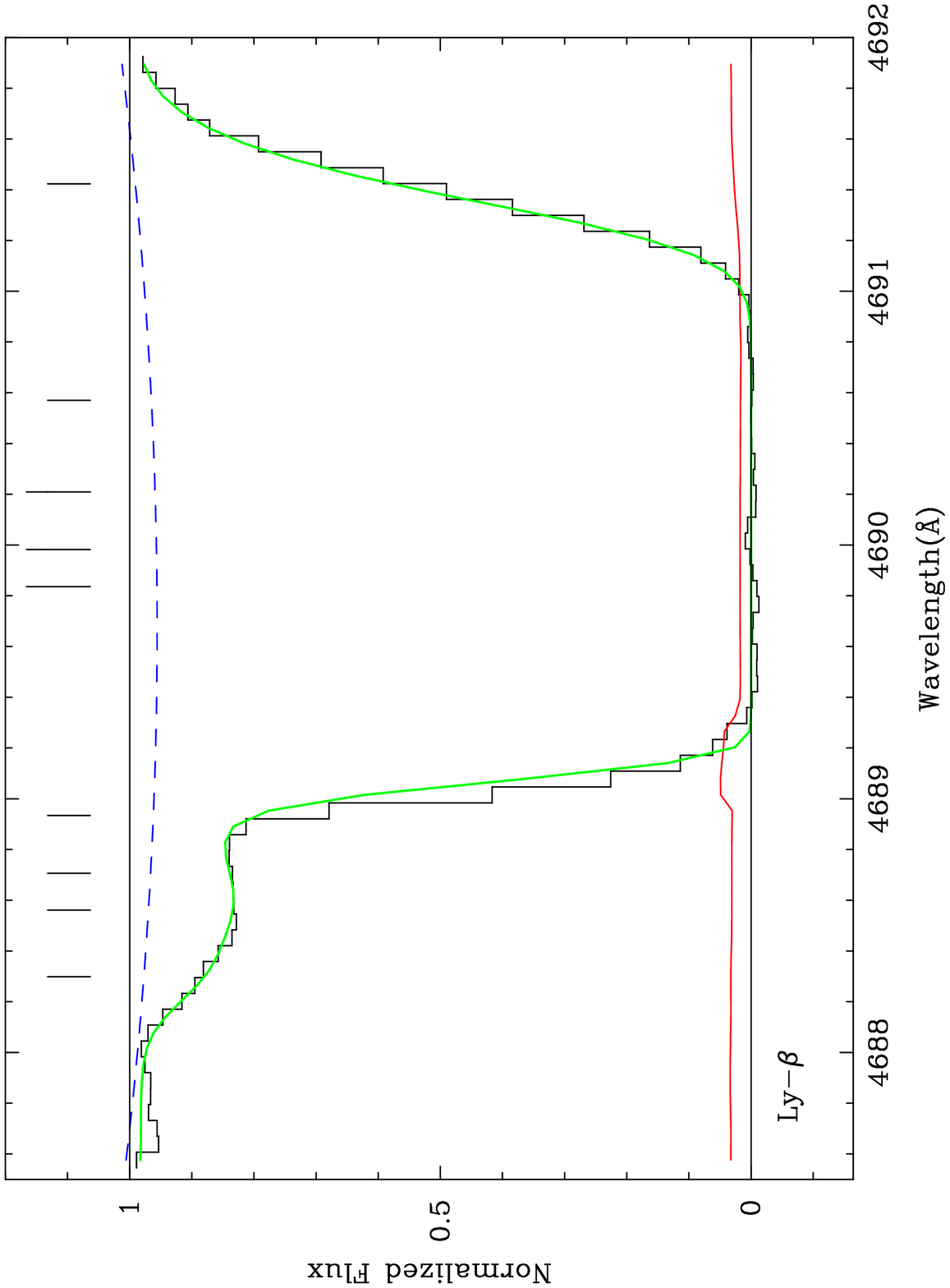,height=\textheight}}
\end{figure}

\end{document}